\documentclass[seceq]{ptptex}

\newcommand{\pMFN}{(\pi M F_N)}
\newcommand{\pMFt}{(\pi M F_\theta)}
\newcommand{\pMft}{(\pi M f)}
\newcommand{\pMFp}{(\pi M F_\varphi)}

\title{%
Adiabatic Evolution of Three `Constants' of Motion \\
for Greatly Inclined Orbits in Kerr Spacetime
}

\author{%       %Use \scshape  for the family name
Katsuhiko \textsc{Ganz},$^1$
Wataru \textsc{Hikida},$^2$
Hiroyuki \textsc{Nakano},$^3$
Norichika \textsc{Sago}$^4$ and 
Takahiro \textsc{Tanaka}$^1$
}

\inst{%     %Affiliation, neglected when [addenda] or [errata]
$^1$
Department of Physics,~Graduate School of Science, Kyoto University,\\
Kyoto 606-8502,~Japan \\
$^2$
Department of Earth and Space Science,~Graduate School of Science,\\
Osaka University, Toyonaka 560-0043, Japan \\
$^3$Department of Physics and Astronomy, 
and Center for Gravitational Wave Astronomy,
The University of Texas at Brownsville, 
Brownsville, Texas 78520, USA\\
$^4$ School of Mathematics, University of Southampton, \\
Southampton, SO17 1BJ, United Kingdom
}

\abst{%         %this abstract is neglected when [addenda] or [errata]
General orbits of a particle of small mass $\mu$ 
around a Kerr black hole of mass $M$ are 
characterized by three parameters:
the energy, the angular momentum and the Carter constant.
The time-averaged rates of change of the energy and the angular momentum
can be obtained by computing the corresponding fluxes of 
gravitational waves emitted by the particle.  
By contrast, the time-averaged rate of change of the Carter constant 
cannot be expressed as a flux of gravitational waves. 
Recently a method to compute this rate of change was proposed by Mino, 
and we refined it into a simplified form. 
In this paper we further extend our previous 
work to give a new formulation without the aid of 
expansion in terms of a small inclination angle.
}

\begin{document}

\maketitle

%%%%%%%%%%%%%%%%%%%%%%%%%%%%%%%%%%%%%%%%%%%%%%%%%%%%
\section{Introduction}
%%%%%%%%%%%%%%%%%%%%%%%%%%%%%%%%%%%%%%%%%%%%%%%%%%%%

Thanks to recent advances in technology, an era of
gravitational  wave astronomy has almost arrived.
There are already several large-scale laser interferometric
gravitational wave detectors in operation. 
Among them are TAMA300 \cite{TAMA},
LIGO \cite{LIGO}, GEO-600 \cite{GEO} and 
VIRGO \cite{VIRGO}. 
The primary targets of these ground-based detectors 
are inspiralling compact binaries, which are expected
to be detected in the near future.

There are also projects for space-based interferometric detectors. 
LISA is now on its research and development %R \& D 
stage~\cite{LISA},
and there is a future plan called DECIGO/BBO~\cite{DECIGO,BBO}.
These space-based detectors can detect gravitational
waves from solar-mass compact objects orbiting supermassive
black holes. To extract physical information concerning such binary
systems, it is essential to know the theoretical
gravitational waveforms with sufficient accuracy.
The black hole perturbation approach is most suited for
this purpose. In this approach, one considers gravitational
waves emitted by a point particle that represents a
compact object orbiting a black hole, assuming the
mass of the particle $\mu$ is much less than that
of the black hole $M$, i.e. $\mu\ll M$.

To lowest order in the mass ratio $\mu/M$, %!!!
the orbit of the particle is a geodesic on the background
geometry of a black hole. Already at this lowest order, 
combined with the assumption of energy and angular
momentum balance between the emitted gravitational waves
and the orbital motion, this approach has proved to
be very powerful in evaluating general relativistic
corrections to the gravitational waveforms, even for
neutron star-neutron star binaries, for which the
assumption of this approach is maximally violated~\cite{MSSTT}.

However, the deviation from the geodesic cannot be completely 
specified by the rates of change of the energy and angular momentum. 
In order to describe general orbits, we need to know the evolution of 
the third ``constant'' of motion, i.e. the Carter constant $Q$. 
For this purpose, we need to evaluate the gravitational self-force 
acting on the particle directly. 
Here, the gravitational self-force is the force due to the 
metric perturbation caused by the particle itself. 

Roughly speaking, there are two levels in the computation of the
self-force in the linear approximation~\cite{tanaka}. 
The advanced level is the direct computation of the time-dependent
self-force, in which one
computes the self-force without any further approximation.  
The main problem here is that the full (bare) metric perturbation 
diverges at the location of the particle, which is assumed to be 
point-like, and hence so does the self-force. 
About a decade ago, a formal expression for the gravitational self-force 
was found, which contains the expression for the Green function divided 
into two parts; the direct part and the tail part~\cite{reaction,QuiWal}.
Later, this formula was reformulated in a more sophisticated manner 
by Detweiler and Whiting~\cite{Detweiler:2002mi}. 
In this new formulation,
the direct part is replaced with the S part and the tail part with 
the R part. The R part 
%is equivalent to the tail
%part as far as the self-force calculation is concerned, but which 
has the improved property that it is a solution of source-free
linearized Einstein equations. 
Thus, what we have to do to obtain a meaningful self-force 
is to compute the R part of the metric perturbation. 
The ``mode sum'' scheme, 
a practical calculation method for the R part, 
has been developed~\cite{Barack:1999wf,Barack:2001bw,Barack:2001gx,
Barack:2002mh,Barack:2002bt,Barack:2003mh,Mino:2001mq}. 
There are several implementations of this scheme in a scalar toy model:
Ref.~\citen{Burko:1999zy} for a static particle, 
Refs.~\citen{Barack:2000zq} and \citen{Burko:2001kr} for radial infall, 
and 
Refs.~\citen{Burko:2000xx}, \citen{Detweiler:2002gi},
\citen{Diaz-Rivera:2004ik} and \citen{Hikida:2004jw}
for a circular orbit.
Extensions to the gravitational case are made  
in Ref.~\citen{Barack:2002ku} for radial infall, 
and in Ref.~\citen{Barack:2007tm} for a circular orbit.

Despite recent progress in the direct computation of the self-force, 
there seem to remain obstacles in computing it for general orbits on 
the Kerr background. 
Furthermore, all results obtained from the 
computation of the self-force are not equally meaningful.
%one starts to realize that the all information about 
%the self-force are not equally valuable. 
A few time-averaged combinations 
composed of the self-force are important for the prediction of
the gravitational waveform. The most important quantities are the 
time-averaged rates of change of the ``constants'' of motion. 
Here a constant of motion means the quantity, such as the energy or
angular momentum, that stays constant for background geodesics. 
Hence, the other, easier level of computing the self-force is 
to compute these rates of change of the ``constants'' of motion. 
Many years ago, Gal'tsov~\cite{Gal'tsov82}
advocated using the radiative part of the metric perturbation to
calculate $dE/dt$ and $dL/dt$. 
The radiative field is defined as half the retarded field minus half
the advanced field. The advantage of using the radiative field is that this
force is relatively easy to compute and is free from divergence. 
Hence we do not have to worry about how
to subtract out the divergent part. 
Gal'tsov showed that
the calculation using the radiative field correctly
reproduces the results obtained by using the balance argument for $dE/dt$
and $dL/dt$ when they are averaged over an infinitely long time
interval. 
Recently, Mino directly verified the validity of applying the same
scheme to the computation 
of the time averaged rates of change of the constants of motion 
including the Carter constant~\cite{Mino:2003yg}.

{}From the equivalence of the two calculations 
for $dE/dt$ and $dL/dt$ shown by Gal'tsov,  
we see that, in order to obtained time-averaged rates of change of
the energy and angular momentum, all we have to do is to compute the 
fluxes of the energy and angular momentum evaluated at infinity and on 
the black hole horizon
%generated by a compact object on a geodesic orbit, 
by using the Teukolsky formalism. 
Therefore, it is expected that the same might be true for the 
Carter constant.
Recently, we have shown that this is indeed the case, obtaining 
a new simplified formula of the time-averaged rate of change for 
$dQ/dt$ written in terms of the asymptotic amplitude of 
gravitational waves~\cite{Sago:2005gd}. This formula requires 
some knowledge of the particle orbit. In this sense,
the rate of change is not expressed as a flux determined by 
the asymptotic form of gravitational waves. 
In our previous paper, using this new formula, we gave 
analytic expressions for the time-averaged 
rates of change of the energy, the angular momentum and the Carter 
constant in the post-Newtonian expansion~\cite{Sago:2006}. 
There, we also made use of the expansion in terms of the
orbital inclination angle for a technical reason. 
Here, we extend our previous results, eliminating the limitation 
to small inclination angles. 

This paper is organized as follows. In \S\ref{GW flux}
we review the basic formalism about how to analytically compute the 
rates of change of the constants of motion due to gravitational 
wave emission for a particle orbiting a Kerr black hole, 
which is developed in Ref.~\citen{Sago:2006}.
The formulas for the rates of change are written in terms of 
the amplitude of each partial wave of the emitted gravitational 
waves. In \S\ref{amplitude} we explain our new method 
for the analytic evaluation of this amplitude for a general orbit with a 
large inclination angle.  
In \S\ref{change rate} substituting the results 
obtained in the preceding section into the formulas 
described in \S\ref{GW flux}, 
we compute the time-averaged rates of change,
$dE/dt$, $dL/dt$, and $dQ/dt$. 
We also compute the phase evolution of gravitational waves.
In \S\ref{summary} we summarize this paper.

%%%%%%%%%%%%%%%%%%%%%%%%%%%%%%%%%%%%%%%%%%%%%%%%%%%%%%%%%%%%%%%%%%%%%%
\section{Basic formulation for adiabatic radiation reaction} 
\label{GW flux}
%%%%%%%%%%%%%%%%%%%%%%%%%%%%%%%%%%%%%%%%%%%%%%%%%%%%%%%%%%%%%%%%%%%%%%

%%%%%%%%%%%%%%%%%%%%%%%%%%%%%%%%%%%%%%%%%%%%%%%%%%%%%%%%%%%%%%%%%%%%%%
%\subsection{Teukolsky formalism} \label{Teukolsky}
%%%%%%%%%%%%%%%%%%%%%%%%%%%%%%%%%%%%%%%%%%%%%%%%%%%%%%%%%%%%%%%%%%%%%%
In this section we give a brief review on the Teukolsky 
formalism~\cite{Teukolsky, oldsupple} as well as the basic formulas
obtained in Ref.~\citen{Sago:2005gd}. 
We consider the background Kerr spacetime in the Boyer-Lindquist
coordinates:
\begin{eqnarray}
ds^2 &=&
-\left(1-\frac{2Mr}{\Sigma}\right)dt^2
-\frac{4Mar\sin^2\theta}{\Sigma}dtd\varphi
+\frac{\Sigma}{\Delta}dr^2
\nonumber \\ &&
\hspace*{2cm} +\Sigma d\theta^2
+\left(r^2+a^2+\frac{2Ma^2r}{\Sigma}\sin^2\theta\right)
\sin^2\theta d\varphi^2, 
\end{eqnarray}
%\begin{eqnarray}
%ds^2=-\left(1-\frac{2Mr}{\Sigma}\right)dt^2
%-\frac{4Mar\sin^2\theta}{\Sigma}\,dt\, d\varphi
%+\frac{\Sigma}{\Delta}dr^2+\Sigma\, d\theta^2
%+\left(r^2+a^2+\frac{2Ma^2r\sin^2\theta}{\Sigma}\right)
%\sin^2\theta\, d\varphi^2, 
%\end{eqnarray}
where
\begin{eqnarray}
\Sigma\equiv r^2+a^2\cos^2\theta,\qquad \Delta\equiv r^2-2Mr+a^2.
\end{eqnarray}
Here, $M$ and $a$ are the mass and the angular momentum of
the black hole, respectively. 
In the Teukolsky formalism, the gravitational perturbation of a Kerr
black hole is described by a master variable $\psi$, which 
satisfies the master equation
\begin{eqnarray}
 {}_{s}\mathcal{O}\psi_{s} = 4\pi \Sigma_{s} T,\label{eq:master}
\end{eqnarray}
where
\begin{eqnarray}
 {}_{s}\mathcal{O} &\equiv & -\left[\frac{(r^2+a^2)^2}{\Delta}
-a^2\sin^2\theta\right]\partial_{t}^2
-\frac{4Mar}{\Delta}\partial_{t}\partial_{\varphi}
-\left[\frac{a^2}{\Delta}-\frac{1}{\sin^2\theta}\right]\partial_{\varphi}^2
\cr &&
+\Delta^{-s}\partial_{r}(\Delta^{s+1}\partial_r)
+\frac{1}{\sin\theta}\partial_{\theta}(\sin\theta\partial_{\theta})
+2s\left[\frac{a(r-M)}{\Delta}+\frac{i\cos\theta}{\sin^2\theta}\right]
\cr &&
\partial_{\varphi}+2s\left[\frac{M(r^2-a^2)}{\Delta}-r-ia\cos\theta\right]
\partial_{t}-s(s\cot^2\theta-1),
\end{eqnarray}
and ${}_sT$ is the source term. 
The master variable $\psi$ is equal to 
$\psi_0$ for $s=2$ and $(r-ia\cos\theta)^4\psi_4$ for
$s=-2$, where 
\begin{eqnarray}
 \psi_0 \equiv -  C_{\alpha\beta\gamma\delta}
  \ell^{\alpha}m^{\beta}\ell^{\gamma}m^{\delta}
,\qquad
 \psi_4 \equiv -  C_{\alpha\beta\gamma\delta}
 n^{\alpha}\overline{m}^{\beta}n^{\delta}\overline{m}^{\delta}
\end{eqnarray}
are the so-called Newman-Penrose quantities.  
Here, $C_{\alpha\beta\gamma\delta}$ is the Weyl tensor, and 
the null vectors $\ell,n,m$ are defined by 
\begin{eqnarray}
\ell^\mu &\equiv& \left((r^2+a^2),\Delta,0,a\right)/\Delta,\
n^\mu \equiv \left((r^2+a^2),-\Delta,0,a\right)/(2\Sigma),
\cr
m^\mu  & \equiv & \left(ia\sin\theta,0,1,i/\sin\theta\right)
/(\sqrt{2}(r+ia\cos\theta)).   
\end{eqnarray}
The bar denotes complex conjugation. 

The master equation (\ref{eq:master}) can be solved 
by decomposing the master variable $\psi$ as
\begin{eqnarray}
 \psi = \sum_{\Lambda} \int d\omega {}_sR_{\Lambda}(r)
{}_{s}S_{\Lambda}(\theta)e^{im\varphi}e^{-i\omega t},
\end{eqnarray}
where $\Lambda$ represents a set of separation constants, $\{\ell,m,\omega\}$.
The equations for the radial and angular parts can be separated, and we obtain
\begin{eqnarray}
&& \left[\Delta^{-s}\frac{d}{dr}\left(\Delta^{s+1}\frac{d}{dr}\right)
+\left(\frac{K^2-2is(r-M)K}{\Delta}+4is\omega r-\lambda\right)\right]
{}_{s}R_{\Lambda}(r) = {}_{s}T_{\Lambda},\cr
&& \left[\frac{1}{\sin\theta}\frac{d}{d\theta}\left(
\sin\theta\frac{d}{d\theta}\right)-a^2\omega^2\sin^2\theta
\right. \cr && \qquad \qquad \quad \left.
-\frac{(m+s\cos^2\theta)^2}{\sin^2\theta}
-2a\omega s\cos\theta +s +2am\omega +\lambda\right]{}_{s}S_\Lambda(\theta)=0,
\end{eqnarray}
where 
$$
K\equiv (r^2+a^2)\omega - ma,
$$ 
and $\lambda$ is the eigenvalue determined by the equation 
for ${}_{s}S_{\Lambda}$. The angular function ${}_{s}S_{\Lambda}$
is called the spin-weighted spheroidal harmonic,
which is usually normalized as
\begin{eqnarray}
 \int^{\pi}_{0} \left({}_{s}S_{\Lambda}\right)^2
%!!! absolute value is necessary?
  \sin\theta\, d\theta = 1.
\end{eqnarray}

We define two independent homogeneous solutions of the radial Teukolsky
equation:
\begin{eqnarray}
 {}_sR^{in}_{\Lambda} &\to& \left\{
\begin{array}{ll}
B^{\rm{trans}}_{\Lambda}\Delta^{-s} e^{-ikr^*},&\quad\text{for}\ r\to r_{+}, \\
r^{-2s-1}B^{\rm{ref}}_{\Lambda}e^{i\omega
 r^*}+r^{-1}B^{\rm{inc}}_{\Lambda}
e^{-i\omega r^*},&\quad\text{for}\ r\to+\infty, \\
\end{array}
\right.\cr
  {}_sR^{up}_{\Lambda} &\to& \left\{
\begin{array}{ll}
C^{\rm{up}}_{\Lambda}e^{ikr^*}+\Delta^{-s}C^{\rm{ref}}_{\Lambda}e^{-ikr^*},
 &\qquad\qquad
  \text{for}\ r\to r_{+}, \\
r^{-2s-1}C^{\rm{trans}}_{\Lambda}
e^{i\omega r^*},&\qquad\qquad \text{for}\ r\to+\infty. \\
\end{array}
\right.
\end{eqnarray}
Here, $k\equiv \omega - ma/2Mr_+$, and $r^*$ is the tortoise coordinate
defined by
\begin{eqnarray}
 r^{*} \equiv 
%\int \frac{dr^*}{dr} dr= %!!! removed
r+ \frac{2Mr_+}{r_+ - r_-}\ln\frac{r-r_+}{2M}
-\frac{2Mr_-}{r_+-r_-}\ln\frac{r-r_-}{2M},
\end{eqnarray}
with $r_{\pm}\equiv M\pm \sqrt{M^2-a^2}$.
A systematic analytic method to compute homogeneous solutions for 
$_sR_\Lambda$ and $_sS_\Lambda$ in the post-Newtonian expansion has been 
developed in Refs.~\citen{Mano:1996gn} and \citen{Mano:1996vt}.

The explicit form of the source term $T_{\Lambda}$ is given by
\begin{eqnarray}
 T_{\Lambda} = 4\int d\Omega dt \rho^{-5}\overline{\rho}^{-1}
(B'_2+B_2^{'*}) e^{-im\varphi+i\omega t}
\frac{{}_{-2} S_{\Lambda}}{\sqrt{2\pi}},\label{eq:energy-momentum}
\end{eqnarray}
where
\begin{eqnarray}
 B'_2 &=& -\frac{1}{2}\rho^8\overline{\rho}\mathcal{L}_{-1}\left[
\rho^{-4}\mathcal{L}_{0}(\rho^{-2}\overline{\rho}^{-1}T_{nn})\right]
\cr &&
-\frac{1}{2\sqrt{2}}\rho^8\overline{\rho}\Delta^2\mathcal{L}_{-1}
\left[\rho^{-4}\overline{\rho}^2 \mathcal{J}_{+}
(\rho^{-2}\overline{\rho}^{-2}\Delta^{-1}T_{\overline{m}n})\right],\cr
B^{'*}_2 &=&-\frac{1}{4}\rho^8\overline{\rho}\Delta^2\mathcal{J}_{+}
\left[\rho^{-4}\mathcal{J}_+(\rho^{-2}\overline{\rho}T_{\overline{m}\overline{m}})\right]
\cr &&
-\frac{1}{2\sqrt{2\pi}}\rho^8\overline{\rho}\Delta^2\mathcal{J}_{+}
\left[\rho^{-4}\overline{\rho}^2\Delta^{-1}\mathcal{L}_{-1}
(\rho^{-2}\overline{\rho}^{-2}T_{\overline{m}n})\right],
\end{eqnarray}
with
\begin{eqnarray*}
 \rho = (r-ia\cos)^{-1},\quad 
\mathcal{L}_s = \partial_\theta + \frac{m}{\sin\theta}
-a\omega \sin\theta +s\cot\theta,\quad
\mathcal{J}_{+} = \partial_r + \frac{iK}{\Delta}.
\end{eqnarray*}
In the above, $T_{nn},T_{\overline{m}n}$ and
$T_{\overline{m}\overline{m}}$ are the tetrad components of the
energy-momentum tensor. 
Here and hereafter, for simplicity, we restrict our attention 
to the $s=-2$ case, and we omit the subscript $s$. 
% and the bar denotes the complex conjugation.

We solve the radial Teukolsky
equation using the Green function method. A solution of the Teukolsky
equation that is purely out-going at infinity and is purely
in-going on the horizon is given by
\begin{eqnarray}
 R_{\Lambda} = \frac{1}{W_{\Lambda}}\left\{
R^{up}_{\Lambda}\int_{r_+}^{r} dr' R^{in}_{\Lambda}
\,T_{\Lambda}\Delta^{-2}
+
R^{in}_{\Lambda}\int_{r}^{\infty}dr'
R^{up}_{\Lambda}\,T_{\Lambda}\Delta^{-2}\right\}, 
\end{eqnarray}
where the Wronskian $W_{\Lambda}$ is given by
\begin{eqnarray}
 W_{\Lambda} = 2i\omega C^{\rm{trans}}_{\Lambda}B^{inc}_{\rm{\Lambda}}.
\end{eqnarray}
Then, the solution has the following asymptotic behavior near the horizon:
\begin{eqnarray}
 R_{\Lambda}(r\to r_+) \to
\frac{B^{\rm{trans}}\Delta^2e^{-ikr^*}}{2i\omega
C^{\rm{trans}}_{\Lambda}B^{\rm{inc}}_{\Lambda}}
\int_{r_+}^{\infty} dr' R^{up}_{\Lambda} T_{\Lambda} \Delta^{-2}
\equiv Z^{H}_{\Lambda}\Delta^2 e^{-ikr^*}.
\end{eqnarray}
Here we defined $Z_\Lambda^H$, the amplitude of each partial wave 
labelled by $\Lambda$.
In the $r\to\infty$ limit, the solution takes the form
\begin{eqnarray}
\label{eq:infinity}
 R_{\Lambda}(r\to\infty) \to
\frac{r^3e^{i\omega r^*}}{2i\omega B^{\rm{inc}}_{\Lambda}}
\int^{\infty}_{r_+}dr'\frac{T_{\Lambda}R^{in}_{\Lambda}}
{\Delta^2}\equiv Z^{\infty}_{\Lambda} r^3 e^{i\omega r^*}.
\end{eqnarray}
Hereafter, we focus on the gravitational waves emitted to infinity, 
partly because the horizon in-going wave does not give 
the dominant contribution in the present calculation, and partly 
because the extension is rather straightforward. 

%\subsection{Adiabatic reaction to the particle motion}
We consider the motion of a point particle, whose coordinates are 
$z^{\alpha} = (t_z(\tau),$ $r_z(\tau),$ $\theta_z(\tau),$ $\varphi_z(\tau))$.
Here $\tau$ is the proper time along the orbit. For geodesic motion,
there are three constants of motion,
\begin{eqnarray}
 \mathcal{E} &\equiv& -u^{\alpha}\xi^{(t)}_{\alpha}
=\left(1-\frac{2Mr_z}{\Sigma}\right)u^t
+\frac{2Mar_z\sin^2\theta_z}{\Sigma}u^{\varphi},\cr
\mathcal{L} &\equiv& u^{\alpha}\xi^{(\varphi)}_{\alpha}
=-\frac{2Mar_z\sin^2\theta_z}{\Sigma}u^t
+\frac{(r_z^2+a^2)^2-\Delta a^2\sin^2\theta_z}{\Sigma}\sin^2\theta_z
u^{\varphi},\cr
\mathcal{Q} &\equiv& K_{\alpha\beta}u^\alpha u^\beta
=\frac{(\mathcal{L}-a\mathcal{E}\sin^2\theta_z)^2}{\sin^2\theta_z}
+a^2\cos^2\theta_z+\Sigma^2(u^\theta)^2,
\end{eqnarray}
where $u^\alpha \equiv dz^\alpha /d\tau$ is 
the four velocity, and $K_{\alpha\beta}$
is the Killing tensor defined by
\begin{eqnarray}
 K_{\alpha\beta}\equiv 2\Sigma \ell_{(\alpha} n_{\beta)}+r^2g_{\alpha\beta}.
\end{eqnarray}
We often use alternative notation for the Carter constant,
$\mathcal{C}\equiv \mathcal{Q}-(a\mathcal{E}-\mathcal{L})^2$. For orbits
on the equatorial plane, $\mathcal{C}$ vanishes. Using these constants of
motion and a new parameter along the trajectory $\lambda$ defined by 
$d\lambda = d\tau/\Sigma$, the geodesic equations become
\begin{eqnarray}
&& \left(\frac{dr_z}{d\lambda}\right)^2 = R(r_z),\qquad
\frac{dt_z}{d\lambda} = -a(a\mathcal{E}\sin^2\theta_z-\mathcal{L})
+\frac{r_z^2+a^2}{\Delta}P(r_z),\cr
&& 
\left(\frac{d\cos\theta_z}{d\lambda}\right)^2 =\Theta(\cos\theta_z),
\qquad
\frac{d\varphi_z}{d\lambda} = -a\mathcal{E} +
\frac{\mathcal{L}}{\sin^2\theta_z}+\frac{a}{\Delta}P(r_z),
\label{eq:geodesic}
\end{eqnarray}
where
\begin{eqnarray}
&& P(r) = \mathcal{E}(r^2+a^2)-a\mathcal{L},\qquad
 R(r) = [P(r)]^2 - \Delta[r^2+\mathcal{Q}],\cr
&& \Theta(\cos\theta) = \mathcal{C}-(\mathcal{C}
 +a^2(1-\mathcal{E}^2)+\mathcal{L}^2)\cos^2\theta
+a^2(1-\mathcal{E}^2)\cos^4\theta.
\end{eqnarray}
It should be noted that the equations for the $r$-component and 
the $\theta$-component are decoupled when they are written 
in terms of $\lambda$. Both $R$
and $\Theta$ are quartic functions of their arguments. Hence, both
solutions are given by elliptic functions. Taking the amplitude of the
radial oscillation as a small parameter, we can systematically expand
the radial solution $r_z(\lambda)$ as a Fourier series. 
For the motion in the
$\theta$-direction, perturbative solutions in Fourier series can be 
systematically obtained even for a large inclination angle 
by taking the coefficient of the
quartic term $a^2(1-\mathcal{E}^2)$ as a small parameter. 
As we explain below,
this expansion is a part of the post-Newtonian expansion. 
We stress that it
is not necessary to restrict the amplitude of oscillation,
$|\theta-\pi/2|$, to small values in this expansion.

The other two equations in~(\ref{eq:geodesic}) are integrated as
\begin{eqnarray}
&& t_z(\lambda) = t^{(r)}(\lambda) + t^{(\theta)}(\lambda) 
+ \left<\frac{dt_z}{d\lambda}\right>_{\!\!\lambda} 
\lambda,\cr &&
 \varphi_z(\lambda) = \varphi^{(r)}(\lambda) +
\varphi^{(\theta)}(\lambda)
+ \left<\frac{d\varphi_z}{d\lambda}\right>_{\!\!\lambda}  \lambda,
\end{eqnarray}
where $\langle\cdots\rangle_\lambda\equiv 
\lim_{\Delta\lambda\to\infty}
 (2\Delta\lambda)^{-1}
\int_{-\Delta\lambda}^{\Delta\lambda}d\lambda \cdots$
represents the time average along the geodesic, and
$t^{(r)}$ and $t^{(\theta)}$, which are defined by
\begin{eqnarray}
 t^{(r)}(\lambda) &\equiv& \int d\lambda \left\{
\frac{(r_z^2+a^2)P(r_z)}{\Delta}-\left<
\frac{(r_z^2+a^2)P(r_z)}{\Delta}\right>_{\!\!\lambda} \right\},\cr
 t^{(\theta)}(\lambda) &\equiv&
-\int d\lambda \left\{
a^2\mathcal{E}\sin^2\theta_z-\left<
a^2\mathcal{E}\sin^2\theta_z\right>_{\lambda} 
\right\}
\end{eqnarray}
are periodic functions of periods $2\pi/\Omega_r$ and
$2\pi/\Omega_\theta$, respectively. 
The functions $\varphi^{(r)}$ and
$\varphi^{(\theta)}$ are defined in the same way,
\begin{eqnarray}
 \varphi^{(r)}(\lambda) &\equiv& \int d\lambda \left\{
\frac{aP(r_z)}{\Delta}-\left<
\frac{aP(r_z)}{\Delta}\right>_{\!\!\lambda}
\right\},\cr
 \varphi^{(\theta)}(\lambda) &\equiv&
\int d\lambda \left\{
\frac{\cal{L}}{\sin^2\theta_z}-\left<
\frac{\cal{L}}{\sin^2\theta_z}
\right>_{\lambda} 
\right\}.
\end{eqnarray}

Using Eqs.~(\ref{eq:geodesic}), the tetrad components of the energy
momentum tensor are expressed as
\begin{eqnarray}
 T_{nn} &=& \mu \frac{C_{nn}}{\sin\theta}
\delta(r-r(t))\delta(\theta-\theta(t))\delta(\varphi-\varphi(t)),\cr
 T_{\overline{m}n} &=& \mu \frac{C_{\overline{m}n}}{\sin\theta}
\delta(r-r(t))\delta(\theta-\theta(t))\delta(\varphi-\varphi(t)),\cr
 T_{\overline{m}\overline{m}} &=& \mu
 \frac{C_{\overline{m}\overline{m}}}
{\sin\theta}
\delta(r-r(t))\delta(\theta-\theta(t))\delta(\varphi-\varphi(t)),
\end{eqnarray}
where
\begin{eqnarray}
 C_{nn} &\equiv & 
\frac{\rho^2\overline{\rho}^2}{4\dot{t}}
\left[\mathcal{E}(r^2+a^2)-a\mathcal{L}+\frac{dr}{d\lambda}\right]^2,\cr
 C_{\overline{m}n} &\equiv & -\frac{\rho^2\overline{\rho}}{2\sqrt{2}\dot{t}}
\left[\mathcal{E}(r^2+a^2)-a\mathcal{L}+\frac{dr}{d\lambda}\right]
\left[i\sin\theta\left(a\mathcal{E}-\frac{\mathcal{L}}{\sin^2\theta}\right)
-\frac{1}{\sin\theta}\frac{d\cos\theta}{d\lambda}\right],\cr
 C_{\overline{m}\overline{m}} &\equiv & 
\frac{\rho^2}{2\dot{t}}\left[i\sin\theta\left(a\mathcal{E}-\frac{\mathcal{L}}{\sin^2\theta}\right)
-\frac{1}{\sin\theta}\frac{d\cos\theta}{d\lambda}\right]^2,
\end{eqnarray}
and $\dot{t}\equiv dt/d\lambda$. After some calculation,
Eq.~(\ref{eq:energy-momentum}) finally becomes
\begin{eqnarray}
 T_{\Lambda} &=& \mu \int dt\, e^{i\omega t-im\varphi(t)} \Delta^2\left[
(A_{nn0}+A_{\bar{m}n0}+A_{\bar{m}\bar{m}0})\delta(r-r(t))
\right. \cr && \left.\qquad
+\left\{(A_{\bar{m}n1}+A_{\bar{m}\bar{m}1})\delta(r-r(t))\right\}_{,r}
+\left\{(A_{\bar{m}\bar{m}2})\delta(r-r(t))\right\}_{,rr}
\right],\label{eq:energy-momentum2}
\end{eqnarray}
where 
$$
\mathcal{L}^{\dagger}_\sigma 
  \equiv \partial_\theta - \frac{m}{\sin\theta}
  +a\omega \sin\theta +\sigma\cot\theta,
$$ 
and  
\begin{eqnarray}
A_{nn0}&=& \frac{-2}{\sqrt{2\pi}\Delta^2}
C_{nn}\rho^{-2}\bar{\rho}^{-1}\mathcal{L}_{1}^{\dagger}
\left\{\rho^{-4}\mathcal{L}_2^{\dagger}(\rho^3 S_\Lambda)\right\},\cr
A_{\bar{m}n0}&=& \frac{2}{\sqrt{\pi}\Delta}
C_{\bar{m}n}\rho^{-3}\left[(\mathcal{L}^{\dagger}_2 S_\Lambda)
\left(\frac{iK}{\Delta}+\rho+\bar{\rho}\right)
-a\sin\theta S_\Lambda\frac{K}{\Delta}(\bar{\rho}-\rho)\right],\cr
A_{\bar{m}\bar{m}0} &=& -\frac{1}{\sqrt{2\pi}}\rho^{-3}\bar{\rho}
C_{\bar{m}\bar{m}}S_\Lambda\left[-i\left(\frac{K}{\Delta}\right)_{,r}
-\frac{K^2}{\Delta^2}+2i\rho\frac{K}{\Delta}\right],\cr
A_{\bar{m}n1}&=& \frac{2}{\sqrt{\pi}\Delta}\rho^{-3}
C_{\bar{m}n}\left[\mathcal{L}^{\dagger}_2S_\Lambda
+ia\sin\theta(\bar{\rho}-\rho)S_\Lambda\right],\cr
A_{\bar{m}\bar{m}1}&=& -\frac{2}{\sqrt{2\pi}}\rho^{-3}\bar{\rho}
C_{\bar{m}\bar{m}}S_\Lambda\left(i\frac{K}{\Delta}+\rho\right),\cr
A_{\bar{m}\bar{m}2}&=&-\frac{1}{\sqrt{2\pi}}\rho^{-3}\bar{\rho}
C_{\bar{m}\bar{m}}S_\Lambda .
\end{eqnarray}
Inserting Eq.~(\ref{eq:energy-momentum2}) into Eq.~(\ref{eq:infinity}),
we obtain $Z_{\Lambda}$ as
\begin{eqnarray}\label{eq:Teukol-formula}
 Z_{\Lambda} = \frac{\mu}{2i\omega B^{\rm{inc}}}
\int_{-\infty}^{\infty} dt e^{i\omega t-im\varphi}&&
\left[
R^{in}_{\Lambda}\{A_{nn0}+A_{\bar{m}n0}+A_{\bar{m}\bar{m}0}\}
\right. \cr && \left.
-\frac{dR^{in}_{\Lambda}}{dr}
\{A_{\bar{m}n1}+A_{\bar{m}\bar{m}1}\}
+\frac{d^2R^{in}_{\Lambda}}{dr^2}
A_{\bar{m}\bar{m}2}
\right]_{r=r(t)}.
\end{eqnarray}
Furthermore, when we consider bound orbits, the frequency spectrum of
$T_{\Lambda}$ becomes discrete. Therefore $Z_{\Lambda}$ takes the form
\begin{eqnarray}
 Z_{\Lambda} = \sum_{n_r,n_\theta}
  2\pi\delta(\omega-\omega^{n_r,n_\theta}_{m})\tilde{Z}_{\tilde{\Lambda}}
\sqrt{4\pi}\omega,
\end{eqnarray}
with
\begin{eqnarray}
 \omega^{n_r,n_\theta}_{m}\equiv
\left<\frac{dt_z}{d\lambda}\right>_{\!\!\lambda}^{-1}\left(
m\left<\frac{d\varphi_z}{d\lambda}\right>_{\!\!\lambda} 
+n_r\Omega_r+n_\theta\Omega_\theta\right).
\end{eqnarray}
Here $\tilde{\Lambda}$ denotes the set of parameters 
$\{\ell,m,n_r,n_\theta\}$. Then the time-averaged luminosity is given by
\begin{eqnarray}
 \left<\frac{d\mathcal{E}}{dt}\right>
    = \mu^{-1}\sum_{\tilde{\Lambda}}
\left|\tilde{Z}_{\tilde{\Lambda}}\right|^2, \label{eq:dEdt}
\end{eqnarray}
and the time-averaged angular momentum flux is given by
\begin{eqnarray}
 \left<\frac{d\mathcal{L}}{dt}\right> = \mu^{-1}
  \sum_{\tilde{\Lambda}}
\frac{m}{\omega^{n_r,n_\theta}_{m}}\left|\tilde{Z}_{\tilde{\Lambda}}\right|^2.
\label{eq:dLdt}
\end{eqnarray}
Here $\langle\cdots\rangle$ represents
the time average with respect to the coordinate time $t$.
We quote the formulas for calculating 
$\left<d\mathcal{C}/dt\right>$ 
from our previous paper~\cite{Sago:2005gd} :
\begin{eqnarray}
\left<\frac{d \mathcal{C}}{dt}\right> &=&  
\left<\frac{d \mathcal{Q}}{dt}\right>
-2\left(a\mathcal{E}-\mathcal{L}\right)
\left(a\left<\frac{d \mathcal{E}}{dt}\right>
-\left<\frac{d \mathcal{L}}{dt}\right>\right),
\cr
 \left<\frac{d\mathcal{Q}}{d t}\right> &=& 
  2\left<\frac{(r^2+a^2)P}{\Delta}\right>_{\!\!\lambda} \left<\frac{d\mathcal{E}}{dt}\right>
 -2\left<\frac{aP}{\Delta}\right>_{\!\!\lambda} 
  \left<\frac{d\mathcal{L}}{dt}\right>
\cr && 
 + \mu^{-1}
  \sum_{\tilde{\Lambda}} \frac{2n_r\Omega_r}{\omega^{n_r,n_\theta}_{m}}
\left|\tilde{Z}_{\tilde{\Lambda}}\right|^2.
\label{eq:dCdt}
\end{eqnarray}
%And the details of calculation of homogeneous solutions and spheroidal
%harmonics are in the Sago's paper.

%%%%%%%%%%%%%%%%%%%%%%%%%%%%%%%%%%%%%%%%%%%%%%%%%%%%%%%%%%%%%%%%%%%%%%
\section{Amplitude of partial wave} 
\label{amplitude}
%%%%%%%%%%%%%%%%%%%%%%%%%%%%%%%%%%%%%%%%%%%%%%%%%%%%%%%%%%%%%%%%%%%%%%
As mentioned above, taking the amplitude of the radial oscillation
and the coefficient of the quartic term of $\Theta(\cos\theta)$, 
which are denoted by ${e}$ and
$\epsilon_0=a^2(1-\mathcal{E}^2)/\mathcal{L}^2$, respectively,
as small parameters, we can systematically expand the 
solutions as Fourier series. Indeed, the solutions are found to take the
forms 
\begin{eqnarray}
r_z(\lambda) = p[1+{e}\cos \Omega_r \lambda]+\mathcal{O}({e}^2),\quad
\cos \theta_z(\lambda) = \sqrt{\frac{y}{y+1}}\sin \Omega_\theta \lambda
+\mathcal{O}(\epsilon_0),
\end{eqnarray}
where $p$ and $y$ are defined in Eqs.~(\ref{rarp}) and (\ref{ydef}). 
There is a difficulty when we actually calculate the amplitude of
a partial wave $\tilde{Z}_{\tilde{\Lambda}}$ using
Eq.~(\ref{eq:Teukol-formula}). 
Here we want to perform the inverse Fourier transformation
analytically. This is possible when $Z_{\Lambda}$ is expanded 
as a power series in sinusoidal functions.  
However, $A_{\bar{m}n}$ and $A_{\bar{m}\bar{m}}$ have
factors of $1/\sin\theta_z$ and $1/\sin^{2}\theta_z$ through 
$C_{\bar{m}n}$ and $C_{\bar{m}\bar{m}}$, respectively.
To make matters worse, $e^{i\varphi}$ is proportional to
$1/\sin\theta$.
This means that 
$\tilde{Z}_{\tilde{\Lambda}}$ has a term proportional to
$\sqrt{(y+1)/(1+y\cos^2\Omega_\theta\lambda)}$, apparently. 
It would seem that this would
prevent us from performing the inverse 
Fourier transformation analytically, unless we expanded this factor 
with respect to $y$, assuming a small inclination angle.
If such problematic terms remain, however, it follows that 
we cannot truncate the series expansion labelled by $n_{\theta}$ 
at a finite order for general orbits.  
This seems quite counter intuitive. 
We expect, therefore, that we can somehow overcome this difficulty. 
In this subsection, we rewrite the source 
term %by using integration by parts 
into a tractable form, which does not contain any inverse power 
of $\sin\theta$.  
We present the results up through $O({e}^2,v^{5})$  as a demonstration, 
where $v$ is a typical velocity of the particle defined more precisely
in Eq.~(\ref{vdef}). 

%%%%%%%%%%%%%%%%%%%%%%%%%%%%%%%%%%%%%%%%%%%%%%%%%%%%%%%%%%%%%%%%%%%%%%
\subsection{General geodesic orbits in Kerr spacetime} \label{Orbits}
%%%%%%%%%%%%%%%%%%%%%%%%%%%%%%%%%%%%%%%%%%%%%%%%%%%%%%%%%%%%%%%%%%%%%%
Generic Kerr geodesics can be specified by fixing three orbital
elements, say, 
the semi-latus rectum $p$, the eccentricity ${e}$, and a dimensionless
inclination parameter ${y}$. % which is related to inclination angle.
We define $p$ and ${e}$ such that the turning points of the
radial motion, the apastron and periastron, are 
\begin{eqnarray}
 r_{a} =  \frac{p}{1-{e}},\quad r_{p}=\frac{p}{1+{e}},
\label{rarp}
\end{eqnarray}
respectively. We also define ${y}$ by
\begin{equation}
  y\equiv \mathcal{C}/\mathcal{L}^2,
\label{ydef}
\end{equation}
which is something like the squared tangent of  
the inclination angle. This parametrization of orbits is 
useful for obtaining an intuitive understanding, but it is, of course, 
equivalent to specifying the three ``constants'' of motion 
$\mathcal{E}$, $\mathcal{L}$ and $\mathcal{C}$.  
Solving the above defining equations, 
$\mathcal{E}$ and $\mathcal{L}$ can be expressed in terms of
$p$, $e$ and $y$ as  
\begin{eqnarray}
 \mathcal{E}&=&1-\frac{1}{2}v^2+\frac{3}{8}v^4-{q}{Y}v^5
   +{e}^2\left\{\frac{1}{2}v^2-\frac{3}{4}v^4+{2q}{Y}v^5\right\},\cr
 \frac{\mathcal{L}}{p}
&=& {v}{Y}+\frac{3Y}{2}v^3-{3q}{Y^2}v^4
+\left({q^2}{Y^3}+\frac{27Y}{8}\right)v^5
\cr &&
+{e}^2\left\{\frac{Y}{2}v^3-{q}{Y^2}v^4
+\left({q^2}{Y^3}+\frac{9Y}{4}\right)v^5\right\},\label{eq:kekka_EL}
\end{eqnarray}
where 
\begin{equation}
v\equiv\sqrt{M/p},
\label{vdef}
\end{equation}
 and 
\begin{equation}
Y\equiv {1\over \sqrt{y+1}} ={{\cal L}\over \sqrt{{\cal C}+{\cal L}^2}}.
\label{Ydef}
\end{equation}
Then, expanding the periodic parts with the period of the
radial oscillations in the geodesic equations in  
powers of $v$ and $e$, the solutions up through $O({e}^2,v^5)$ 
are found to be given by 
\begin{eqnarray}
 r(\lambda) &=& p\sum_{n_r=0}^{n_{\rm{max}}}
\alpha_{n_r}\cos n_r \Omega_r \lambda,\cr
 t^{(r)}(\lambda) &=&
\sum_{n_r=0}^{n_{\rm{max}}} t^{(r)}_{n_r}\sin n_r\Omega_r\lambda,\cr
 \varphi^{(r)}(\lambda) &=&
\sum_{n_r=0}^{n_{\rm{max}}} \varphi^{(r)}_{n_r}\sin n_r\Omega_r\lambda,
\end{eqnarray}
where $n_{\rm{max}}=2$, which is the truncated order in ${e}$, and 
\begin{eqnarray}
\alpha_{0} &=& 1+ {e}^2\left\{\frac{1}{2}-\frac{1}{2}v^2+{q}{Y}v^3
-\left(3-\frac{(1-2Y^2)\,q^2}{2}\right)v^4+{10\,q}{Y}v^5
\right\},\cr
\alpha_{1} &=& {e},\cr
\alpha_{2} &=& {e}^2\left\{
\frac{1}{2}+\frac{1}{2}v^2-{q}{Y}v^3
+\left(3-\frac{(1-2Y^2)\,q^2}{2}\right)v^4-{10\,q}{Y}v^5
\right\},\cr
t^{(r)}_0 &=& 0,\cr\cr
t^{(r)}_1 &=& {e}\left\{
2+4v^2-{6\,q}{Y}v^3
+\left(17-{(1-4Y^2)\,q^2}\right)v^4-{54\,q}{Y}v^5
\right\},\cr\cr
t^{(r)}_2 &=& {e}^2\left\{
\frac{3}{4}+\frac{7}{4}v^2-\frac{13\,qY}{4}v^3
+\left(\frac{81}{8}+\frac{(20Y^2-7)\,q^2}{8}\right)v^4
-\frac{135\,qY}{4}v^5
\right\},\cr
\varphi^{(r)}_0 &=& 0,\cr
\varphi^{(r)}_1 &=& {e}\left\{
-2qv^3+{2\,q^2}{Y}v^4-10\,qv^5\right\},\cr
\varphi^{(r)}_2 &=& {e}^2\left\{
-\frac{q^2Y}{4}v^4+\frac{q}{2}v^5
\right\}. 
\end{eqnarray}
The angular velocity $\Omega_r$ is obtained simultaneously
when we solve $r(\lambda)$, but we give it below, together
with the angular velocity of the $\theta$-oscillations.

The other parts of %!!! ``of'' was added
the geodesic equations which are periodic with the period of 
the $\theta$-oscillations are expanded 
in powers of 
$\epsilon_0\equiv a^2(1-\mathcal{E}^2)/\mathcal{L}^2=\mathcal{O}(v^4)$
and obtained as
\begin{eqnarray}
 \cos\theta(\lambda) &=& {\sqrt{1-Y^2}}
  \sum_{n_\theta=0}^{n_{\rm{max}}} \beta_{n_\theta}
  \sin n_\theta \Omega_\theta \lambda,\cr
 t^{(\theta)}(\lambda) &=& -\frac{p\,q^2 v^4\mathcal{E}}{\mathcal{L}}
\sum_{n_\theta=0}^{n_{\rm{max}}} \hat t^{(\theta)}_{n_\theta}
\sin n_\theta\Omega_\theta\lambda, \quad
\end{eqnarray}
%!!! =0 was added.
where $n_{\rm{max}}=2\times1+2=4$, which is twice the 
truncated order in $\epsilon_0$ plus two,
\begin{eqnarray}
  \beta_0 &=& 0,\quad
  \beta_1 = 1+\epsilon_0
\frac{(Y^2-9Y^4)}{16},\quad
  \beta_2 = 0,\quad
  \beta_3 = \epsilon_0 \frac{(Y^2-Y^4)}{16},\quad
  \beta_4 = 0,\cr
  \hat t^{(\theta)}_0 &=& 0,\quad
  \hat t^{(\theta)}_1 = 0,\quad
  \hat t^{(\theta)}_2 = {(Y-Y^3)}
+\epsilon_0\frac{Y^3(1-7Y^2)(1-Y^2)}{16}
,\cr 
&&  \hat t^{(\theta)}_3 = 0,\quad
  \hat t^{(\theta)}_4 = \epsilon_0 \frac{Y^3(1-Y^2)^2}{64}. 
\end{eqnarray}
For $\varphi^{(\theta)}$, 
we introduce the variable $X=\sin\theta\, e^{i\varphi^{(\theta)}}$. 
Then, it can be expanded as   
\begin{eqnarray}
 \Re(X) &=& \sum_{n_\theta}^{n_{\rm{max}}}
X^{\Re}_{n_\theta}\cos n_\theta \Omega_\theta \lambda,\quad
 \Im(X) =  \sum_{n_\theta}^{n_{\rm{max}}}
X^{\Im}_{n_\theta}\sin n_\theta \Omega_\theta \lambda,
\end{eqnarray}
where
\begin{eqnarray}
 X^{\Re}_0 &=&
  \frac{(Y+1)}{2}-\epsilon_0\frac{Y^2(1-9Y)(1-Y^2)}{32},\quad
 X^{\Re}_1 = 0,\cr &&
 X^{\Re}_2 = \frac{(1-Y)}{2}-\epsilon_0\frac{Y^3(1-Y^2)}{4},\quad
X^{\Re}_3 = 0,\quad
 X^{\Re}_4 = \epsilon_0\frac{Y^2(1+Y)(1-Y)^2}{32},\cr
 X^{\Im}_0 &=& 0,\quad
 X^{\Im}_1 = 0,\quad
 X^{\Im}_2 =
   -\frac{(1-Y)}{2}+\epsilon_0\frac{Y^2(1+5Y)(1-Y^2)}{16},\cr
 && X^{\Im}_3 = 0,\quad
 X^{\Im}_4 = -\epsilon_0\frac{Y^2(1+Y)(1-Y)^2}{32}.
\end{eqnarray}
Using Eq.~(\ref{eq:kekka_EL}), we 
re-expand the above solutions 
in powers of $v$ and $e$ to obtain 
\begin{eqnarray}
 \cos\theta(\lambda) &=& {\sqrt{1-Y^2}}
  \sum_{n_\theta=0}^{n_{\rm{max}}}\beta_{n_\theta}
  \sin n_\theta \Omega_\theta \lambda,\qquad
 t^{(\theta)}(\lambda) = 
\sum_{n_\theta=0}^{n_{\rm{max}}} t^{(\theta)}_{n_\theta}
\sin n_\theta\Omega_\theta\lambda,\cr
 \Re(X) &=& \sum_{n_\theta=0}^{n_{\rm{max}}}
X^{\Re}_{n_\theta}\cos n_\theta \Omega_\theta \lambda,\qquad
 \Im(X) =  \sum_{n_\theta=0}^{n_{\rm{max}}}
X^{\Im}_{n_\theta}\sin n_\theta \Omega_\theta \lambda,
\end{eqnarray}
%!!! =0
where
\begin{eqnarray}
  \beta_0 &=& 0,\quad
  \beta_1 = 1+\frac{(1-9Y^2)q^2}{16}v^4
-{e}^2\left\{\frac{(1-9Y^2)q^2}{16}v^4\right\}
,\quad
  \beta_2 = 0,\cr
&&  \beta_3 = 
\frac{(1-Y^2)q^2}{16}v^4-{e}^2\left\{\frac{(1-Y^2)q^2}{16}v^4\right\}
,\quad
  \beta_4 = 0,\cr
 t^{(\theta)}_0 &=& 0,\quad
  t^{(\theta)}_1 = 0,\quad
  t^{(\theta)}_2 = -\frac{(1-Y^2)q^2}{4}v^3 -\frac{(1-Y^2)q^2}{2}v^5
,\
  t^{(\theta)}_3 = 0,\quad
  t^{(\theta)}_4 = 0,\cr
 X^{\Re}_0 &=&
  \frac{(Y+1)}{2}-\frac{(1-9Y)(1-Y^2)q^2}{32}v^4
+{e}^2\left\{\frac{(1-9Y)(1-Y^2)q^2}{32}v^4\right\}
,\quad
 X^{\Re}_1 = 0,\cr
&&  X^{\Re}_2 = \frac{(1-Y)}{2}-\frac{Y(1-Y^2)q^2}{4}v^4
+{e}^2\left\{\frac{Y(1-Y^2)q^2}{4}v^4\right\}
,\quad
 X^{\Re}_3 = 0,\cr
&& X^{\Re}_4 = \frac{(1+Y)(1-Y)^2q^2}{32}v^4
-{e}^2\left\{\frac{(1+Y)(1-Y)^2q^2}{32}v^4\right\}
,\cr
 X^{\Im}_0 &=& 0,\quad
 X^{\Im}_1 = 0,\cr &&
 X^{\Im}_2 =
   -\frac{(1-Y)}{2}+\frac{(1+5Y)(1-Y^2)q^2}{16}v^4
-{e}^2\left\{\frac{(1+5Y)(1-Y^2)q^2}{16}v^4\right\}
,\cr
 && X^{\Im}_3 = 0,\;
 X^{\Im}_4 = 
-\frac{(1+Y)(1-Y)^2q^2}{32}v^4
+{e}^2\left\{\frac{(1+Y)(1-Y)^2q^2}{32}v^4\right\}.
\end{eqnarray}

Finally, the frequencies of radial and $\theta$-oscillations and 
the non-oscillating parts of $dt/d\lambda$ and $d\varphi/d\lambda$ 
are given by 
\begin{eqnarray}
 \frac{\Omega_r}{p^2} &=& \frac{v^3}{M} \left[
1-\frac{3}{2}v^2 +{3\,q}{Y}v^3 +\left(\frac{(1-4Y^2)q^2}{2}
-\frac{45}{8}\right)v^4 +\frac{33\,qY}{2}v^5 
\right. \cr &&\quad \left.
 +{e}^2\left\{\frac{1}{2}v^2-{q}{Y}v^3+\frac{(1+Y^2)\,q^2}{4}v^4
+{2\,q}{Y}v^5 \right\}\right],\cr
 \frac{\Omega_{\theta}}{p^2} &=& \frac{v^3}{M} \left[ 
1+\frac{3}{2}v^2-{3\,q}{Y}v^3
+\left(-\frac{(-7Y^2+1)q^2}{4}+\frac{27}{8}\right)v^4
-\frac{15\,qY}{2}v^5 \right. \cr && \quad \left.
+{e}^2\left\{\frac{1}{2}v^2-{q}{Y}v^3
+\left(\frac{(1+Y^2)q^2}{4}+\frac{9}{4}\right)v^4
 -{7\,q}{Y}v^5\right\}\right]
,\cr
{1\over p^2} \left<\frac{dt}{d\lambda}\right>_{\!\!\lambda}
&=& 
% \frac{v^3}{M} \left[
1+\frac{3}{2}v^2
+\left(\frac{(1-Y^2)q^2}{2}+\frac{27}{8}\right)v^4
-{3\,q}{Y}v^5 
%\right. 
\cr && \;
%\left.
 +{e}^2\left\{
\frac{3}{2}-\frac{1}{4}v^2+{2\,q}{Y}v^3
+\left({(1-2Y^2)q^2}-\frac{99}{16}\right)v^4
+\frac{43\,qY}{2}v^5\right\}
%\right]
,\cr
{1\over p^2}
\left<\frac{d\varphi}{d\lambda}\right>_{\!\!\lambda} &=& 
 \frac{v^3}{M} \left[
1+\frac{3}{2}v^2+{(2-3Y)q}v^3
\right. \cr&&\; \left.
+\left(-\frac{(1+7Y)(1-Y)q^2}{4}+\frac{27}{8}\right)v^4
+\frac{3(2-5Y)q}{2}v^5 \right. \cr && \; \left.
 +{e}^2\left\{
\frac{1}{2}v^2-{q}{Y}v^3
+\left(\frac{(1+Y^2)q^2}{4}+\frac{9}{4}\right)v^4
+{(4-7Y)q}v^5\right\}\right]\!.
\label{2.49}
\end{eqnarray}

%%%%%%%%%%%%%%%%%%%%%%%%%%%%%%%%%%%%%%%%%%%%%%%%%%%%%%%%%%%%%%%%%%%%%%
\subsection{Reformulation of the source term for the Teukolsky equation}
\label{Amplitude2}
%%%%%%%%%%%%%%%%%%%%%%%%%%%%%%%%%%%%%%%%%%%%%%%%%%%%%%%%%%%%%%%%%%%%%%
Let us now discuss the method for rewriting the source term of 
the Teukolsky equation.
The source term~(\ref{eq:Teukol-formula}) contains inverse 
powers of $\sin\theta$ 
in $C_{\bar{m}\bar{m}}$, $C_{\bar{m}n}$, and $e^{-im\varphi}\sim 
(\sin\theta)^{-|m|}$. These factors prevent us from 
performing the inverse Fourier transformation analytically. 
Therefore we want to remove all 
inverse powers of $\sin\theta$.  
%in order to calculate the adiabatic evolutions of the
%constants of motion. 
Fortunately, the spheroidal harmonics $S_\Lambda$, 
$\mathcal{L}_2^\dagger S_\Lambda$ and 
$\mathcal{L}_1^\dagger \mathcal{L}_2^\dagger S_\Lambda$
are proportional to $(\sin\theta)^{||m|-2|}$,
$(\sin\theta)^{||m|-1|}$ and $(\sin\theta)^{|m|}$, respectively. 
Therefore some of the inverse powers of $\sin\theta$ cancel out, 
but we immediately 
find that some of them remain after this cancellation.
In the following we show how we can eliminate all 
these annoying factors. 

For convenience, we introduce the new angular functions
\begin{eqnarray}
 {}_2\tilde{S}\equiv \frac{S_\Lambda}{(\sin\theta)^{|m|-2}},\quad
 {}_1\tilde{S}\equiv \frac{\mathcal{L}_2^\dagger \, S_\Lambda}
   {(\sin\theta)^{|m|-1}},\quad
 {}_0\tilde{S}\equiv \frac{\mathcal{L}_1^\dagger \mathcal{L}_2^\dagger \, S_\Lambda}
 {(\sin\theta)^{|m|}},
\end{eqnarray}
and 
\begin{eqnarray}
 {}_\sigma \Xi_m &\equiv& {}_\sigma\tilde{S}X_m, \quad 
 X_m=(\sin\theta)^{|m|} e^{-im\varphi^{(\theta)}}=
 \left\{ \begin{array}{ll} \bar{X}^{m},& (m>0)\\ 
1,& (m=0)
 \\ X^{|m|}.& (m<0) \\
 \end{array} \right. 
\end{eqnarray}
If they are truncated at a finite post-Newtonian order,  
these new functions ${}_\sigma\tilde{S}$, and 
hence ${}_\sigma\Xi_m$, 
have no inverse powers of $\sin\theta$, being expressed as 
polynomials of $\cos\theta$.  
There are a few special cases in which ${}_\sigma\tilde{S}$ 
contains an additional overall factor of $\sin\theta$. 
We have 
${}_2\tilde{S}=\sin^2\theta
(1\pm\cos\theta)\,\times$(a polynomial of $\cos\theta$)
for $m=\pm 1$, 
while 
${}_1\tilde{S}=\sin^2\theta\times$(polynomial of $\cos\theta$), and
${}_2\tilde{S}=\sin^4\theta\times$(polynomial of $\cos\theta$)
for $m=0$. 
In addition, $X_m$ can be expressed as a Fourier series of
$\exp(i\Omega_\theta\lambda)$, because it contains a positive power 
of $X$ or $\bar{X}$. 
Using $_\sigma\Xi_m$, the amplitude of the partial wave is 
rewritten as 
\begin{eqnarray}
 Z_{\Lambda}&=&\frac{\mu}{2i\omega B^{\rm{inc}}}
 \int_{-\infty}^{\infty} d\lambda e^{i\omega t-im(\varphi-\varphi^{(\theta)})}
 \left( \frac{-1}{2\sqrt{2\pi}}\frac{\bar{\rho}}{\rho} \right)
 \Bigg[ \frac{D_r^2}{\Delta^2}
 \left({}_0\Xi_m -2ia\rho \, {}_1\Xi_m \right) R_{\Lambda} \cr
 &&\hspace*{15mm}-\frac{2D_r D_\theta}{\Delta}
 \left({}_1\Xi_m(\mathcal{J}_{-}-(\rho+\bar{\rho}))
 +ia\, {}_2\Xi_m (\bar{\rho}-\rho)\mathcal{J}_{-}\right) R_{\Lambda}
\cr &&\hspace*{60mm}
 +D_\theta^2 \, {}_2\Xi_m (\mathcal{J}_{-}^{2}-2\rho\mathcal{J}_{-})
 R_{\Lambda} \Bigg]_{r=r(\lambda)},
 \label{reformalized Base}
\end{eqnarray}
where we have defined 
\begin{eqnarray}
 \mathcal{J_{-}} & \equiv & \partial_r-iK/\Delta, \cr
 D_r& \equiv &\mathcal{E}(r^2+a^2)-a\mathcal{L}+\frac{dr}{d\lambda}, \cr
 D_\theta&\equiv 
  &i\left(a\mathcal{E}-\frac{\mathcal{L}}{\sin^2\theta}\right)
 -\frac{1}{\sin^2\theta}\frac{d\cos\theta}{d\lambda}.
\end{eqnarray}
In this expression, it is clear that 
inverse powers of $\sin\theta$ are contained only through $D_\theta$, 
which contains the factor $(\sin\theta)^{-2}$, 
and therefore the index of the inverse powers of $\sin\theta$ 
is at most four. We emphasize that this index 
is less for $|m|=0$ or $1$, 
since ${}_2\Xi_m$ and/or ${}_1\Xi_m$
contain additional positive powers of $\sin\theta$. 
Thus, we have to treat
three cases, $m=0$, $m=\pm 1$, and $|m|\ge 2$, separately.
Here we discuss only the case with $|m|\ge 2$, 
and the other simpler cases with $|m|=0$ or $\pm 1$ 
are deferred to Appendix A.

Equation~(\ref{reformalized Base}) has 
inverse powers of $\sin\theta$ in $D_\theta \, {}_1\Xi_m$,
$D_\theta \, {}_2\Xi_m$, and $D_\theta^2 \, {}_2\Xi_m$. 
First we give a prescription for the terms containing
the factor $(\sin\theta)^{-2}$, i.e.
$D_\theta\, {}_{1}\Xi_m$ and $D_\theta\, {}_{2}\Xi_m$.
After some calculation, we obtain (for $\sigma\ge 1$)
\begin{eqnarray}
 \frac{d}{d\lambda}{}_\sigma\Xi_m &=& -\frac{d\cos\theta}{d\lambda}\,
 {}_{\sigma-1}\Xi_m
 +m\left[i\left(a\mathcal{E}-\frac{\mathcal{L}}{\sin^2\theta}\right)
 -\frac{1}{\sin^2\theta}\frac{d\cos\theta}{d\lambda}\right] 
 {}_\sigma\Xi_m \nonumber \\
 &&{}+\left[a\omega\frac{d\cos\theta}{d\lambda}
 +im\left<\frac{\mathcal{L}}{\sin^2\theta}\right>_{\!\!\lambda} 
 -ima\mathcal{E} \right] {}_\sigma\Xi_m,
\end{eqnarray}
where we have used 
\begin{eqnarray}
 \frac{\mathcal{L}}{\sin^2\theta}=
 \left<\frac{\mathcal{L}}{\sin^2\theta}\right>_{\!\!\lambda} 
 +\frac{d\varphi^{(\theta)}}{d\lambda}.
\end{eqnarray}
Using this relation, we can derive 
\begin{eqnarray}
 D_\theta \, {}_\sigma\Xi_m&=&F_m({}_{\sigma-1}\Xi_m,\,{}_\sigma\Xi_m),
 \label{D_relation} \\
 F_m(A,B)&\equiv&\frac{1}{m}\left[
  \frac{d\cos\theta}{d\lambda}A+
  \left(\frac{d}{d\lambda}-a\omega\frac{d\cos\theta}{d\lambda}
   -im\left<\frac{\mathcal{L}}{\sin^2\theta}\right>_{\!\!\lambda}
    +ima\mathcal{E}
  \right)B
 \right]. \nonumber
\end{eqnarray}
Since there are no inverse powers of $\sin\theta$ in 
$F_m({}_{\sigma-1}\Xi_m,\,{}_{\sigma}\Xi_m)$, 
the inverse Fourier transformation of this term can be easily performed. 
Obviously, the above formula is not valid for the case with $m=0$ 
because $F_m$ contains $m^{-1}$.
In $\left<\mathcal{L}/\sin^2\theta \right>_{\lambda}$ 
it seems that $(\sin\theta)^{-2}$ is remaining, but it is just a constant,
\begin{eqnarray}
\hspace*{-10mm}
 \left<\frac{\mathcal{L}}{\sin^2\theta}\right>_{\!\!\lambda} &=&
  \frac{\Omega_\theta}{2\pi}\int_0^{\frac{2\pi}{\Omega_\theta}}
  \!\!\! d\lambda \, \frac{\mathcal{L}}{\sin^2\theta}
  =\frac{2\Omega_\theta}{\pi}\int_{0}^{(\cos\theta)_{\rm max}} d\cos\theta
  \frac{1}{\sqrt{\Theta(\cos\theta)}}\frac{\mathcal{L}}{1-\cos^2\theta},
\end{eqnarray}
which can be analytically integrated in the post-Newtonian expansion. 

Next, we consider the remaining term,
$D_\theta^2 \, {}_2\Xi_m$. Applying Eq.~(\ref{D_relation}) 
twice, we obtain 
\begin{eqnarray}
 D_\theta^2 \, {}_2\Xi_m
 &=&D_\theta F_m({}_1 \Xi_m,\, {}_2\Xi_m) \nonumber \\
 &=& F_m(D_\theta \,{}_1\Xi_m,\,D_\theta\,{}_2\Xi_m)
  -\frac{1}{m} \frac{dD_\theta}{d\lambda} {}_2\Xi_m \cr
 &=& F_m(F_m({}_0\Xi_m,\,{}_1\Xi_m),
      F_m({}_1\Xi_m, \, {}_2\Xi_m))
  -\frac{1}{m} \frac{dD_\theta}{d\lambda} {}_2\Xi_m.
\label{DD}
\end{eqnarray}
Using the geodesic equation for $\cos\theta$, one can show 
\begin{eqnarray}
 \frac{dD_\theta}{d\lambda}=\cos\theta
 (-D_\theta^2+2ia\mathcal{E}D_\theta+a^2). 
\end{eqnarray}
Substituting this relation into Eq.~$(\ref{DD})$ and 
applying Eq.~(\ref{D_relation}) again, 
we obtain
\begin{eqnarray}
 D_\theta^2 \, {}_2\Xi_m
 &=& \left(1+\frac{\cos\theta}{m}\right)
  \Bigg\{ F_m(F_m({}_0\Xi_m,\,{}_1\Xi_m),
  F_m({}_1\Xi_m,\, {}_2\Xi_m)) \cr
 &&{}-\frac{2ia\mathcal{E} \cos\theta}{m} F_m({}_1\Xi_m,\, {}_2\Xi_m) 
 -\frac{a^2 \cos\theta}{m} \, {}_2\Xi_m\Bigg\} 
 +\frac{1}{m^2}(1-\sin^2\theta)D_\theta^2 \,{}_2\Xi_m.
 \nonumber
\end{eqnarray}
Here, $D_\theta^2\,{}_2\Xi_m$ 
appears also 
in the last term on the right-hand side. 
After solving this equation for $D_\theta^2\,{}_2\Xi_m/m^2$, 
we finally arrive at 
\begin{eqnarray}
 D_\theta^2\,{}_2\Xi_m
 &=& \frac{m^2}{m^2-1}\left(1+\frac{\cos\theta}{m}\right)
  \Bigg\{ F_m(F_m({}_0\Xi_m,\,{}_1\Xi_m),
  F_m({}_1\Xi_m,\,{}_2\Xi_m))\cr
 && {}-\frac{2ia\mathcal{E} \cos\theta}{m} F_m({}_1\Xi_m,\,{}_2\Xi_m)
  -\frac{a^2 \cos\theta}{m} {}_2\Xi_m \Bigg\} 
\cr &&\qquad
  - \frac{1}{m^2-1}(D_\theta \sin^2\theta) F_m({}_1\Xi_m, {}_2\Xi_m),  
\end{eqnarray}
which is free from inverse powers of $\sin\theta$. 
Note that this formula is not valid for $m=0$ and $m=\pm1$
because of the factors of $m^{-1}$ and $(m^2-1)^{-1}$.

Combining the above results, the partial wave amplitude for $|m|\ge 2$ is
expressed as 
\begin{eqnarray}
 Z_{\Lambda}&=&\frac{\mu}{2i\omega B^{\rm{inc}}}
 \int_{-\infty}^{\infty} d\lambda \,
 e^{i\omega t-im(\left<\frac{d\varphi}{d\lambda}\right>_{\!\!\lambda} 
   +\varphi^{(r)})}
 \left( \frac{-1}{2\sqrt{2\pi}}\frac{\bar{\rho}}{\rho} \right)
 \Bigg[ \frac{D_r^2}{\Delta^2}
 \left({}_0\Xi_m -2ia\rho\,{}_1\Xi_m \right) R_{\Lambda} \cr
 &&{}-\frac{2D_r}{\Delta} \left\{ F_m({}_0\Xi_m,\,{}_1\Xi_m)
 (\mathcal{J}_{-}-(\rho+\bar{\rho}))
 +iaF_m({}_1\Xi_m,\,{}_2\Xi_m)(\bar{\rho}-\rho)\mathcal{J}_{-} \right\}
 R_{\Lambda} \cr
 &&{}+ \frac{1}{m^2-1}\Big[(m+\cos\theta)
   \Big\{ m F_m(F_m({}_0\Xi_m,\,{}_1\Xi_m),F_m({}_1\,\Xi_m,\,{}_2\Xi_m))
\cr  &&\hspace*{25mm} {}  
-{2ia\mathcal{E} \cos\theta}\,F_m({}_1\Xi_m,\,{}_2\Xi_m) -{a^2
\cos\theta}\, {}_2\Xi_m \Big\}
\cr &&\hspace*{25mm}
  -(D_\theta \sin^2\theta) F_m({}_1\Xi_m,\,{}_2\Xi_m) \Big]
  (\mathcal{J}_{-}^{2}-2\rho\mathcal{J}_{-})
 R_{\Lambda} \Bigg]_{r=r(\lambda)}.
\label{eq:ZLambda}
\end{eqnarray}
This formula has no $(\sin\theta)^{-1}$ factor in the integrand.
The formulas for $|m|\le 1$ are presented in Appendix~\ref{app:mle1}.

%%%%%%%%%%%%%%%%%%%%%%%%%%%%%%%%%%%%%%%%%%%%%%%%%%%%%%%%%%%%%%%%%%%%%%
%%%%%%%%%%%%%%%%%%%%%%%%%%%%%%%%%%%%%%%%%%%%%%%%%%%%%%%%%%%%%%%%%%%%%%
\section{Results}
\label{change rate}
%%%%%%%%%%%%%%%%%%%%%%%%%%%%%%%%%%%%%%%%%%%%%%%%%%%%%%%%%%%%%%%%%%%%%%
%%%%%%%%%%%%%%%%%%%%%%%%%%%%%%%%%%%%%%%%%%%%%%%%%%%%%%%%%%%%%%%%%%%%%%

%%%%%%%%%%%%%%%%%%%%%%%%%%%%%%%%%%%%%%%%%%%%%%%%%%%%%%%%%%%%%%%%%%%%%%
\subsection{The evolution of orbital parameters}
\label{evolution of constants}
%%%%%%%%%%%%%%%%%%%%%%%%%%%%%%%%%%%%%%%%%%%%%%%%%%%%%%%%%%%%%%%%%%%%%%
Substituting Eq.~(\ref{eq:ZLambda}) into 
Eqs.~(\ref{eq:dEdt}), (\ref{eq:dLdt}) and (\ref{eq:dCdt}),
the time-averaged rates of change for  the three constants
of motion up through $O({e}^2,v^5)$ are given by 

\begin{eqnarray}
&&\left<\frac{d\mathcal{E}}{d t}\right> =
-\frac{32}{5}\left(\frac{\mu}{M^2}\right)v^{10}(1-{e}^2)^{3/2}
\left[
\left(1+\frac{73}{24}{e}^2\right)
-\left(\frac{1247}{336}+\frac{9181}{672}{e}^2\right)v^2
\right. \cr && \left.\hspace*{20mm}
-\left(\frac{73Y}{12}+\frac{823Y}{24}{e}^2\right)q v^3
+\left(4+\frac{1375}{48}{e}^2\right)\pi v^3
\right. \cr && \left.\hspace*{20mm}
-\left(\frac{44711}{9072}+\frac{172157}{2592}{e}^2\right)v^4
\right. \cr && \left.\hspace*{20mm}
-\left(\frac{329}{96}-\frac{527Y^2}{96}
+\left\{\frac{4379}{192}-\frac{6533Y^2}{192}\right\}{e}^2\right)q^2v^4
\right. \cr && \left.\hspace*{20mm}
-\left(\frac{8191}{672}+\frac{44531}{336}{e}^2\right)\pi v^5
+\left(\frac{3749Y}{336}+\frac{1759Y}{56}{e}^2\right)qv^5
\right],\cr
&&\left<\frac{d\mathcal{L}}{d t}\right> =
-\frac{32}{5}\left(\frac{\mu}{M^2}\right)Mv^{7}(1-{e}^2)^{3/2}
\left[
\left(Y+\frac{7Y}{8}{e}^2\right)
-\left(\frac{1247Y}{336}+\frac{425Y}{336}{e}^2\right)v^2
\right.\cr && \left.\hspace*{20mm}
+\left(\frac{61}{24}-\frac{61Y^2}{8} 
+ \left\{\frac{63}{8}-\frac{91Y^2}{4}\right\}{e}^2\right)qv^3
+\left(4Y+\frac{97Y}{8}{e}^2\right)\pi v^3
\right. \cr && \left.\hspace*{20mm}
-\left(\frac{44711Y}{9072}+\frac{302893Y}{6048}{e}^2\right)v^4
\right. \cr && \left.\hspace*{20mm}
-\left(\frac{57Y}{16}-\frac{45Y^3}{8}
+\left\{\frac{201Y}{16}-\frac{37Y^3}{2}\right\}{e}^2\right)q^2 v^4
\right. \cr && \left.\hspace*{20mm}
-\left(\frac{8191Y}{672}+\frac{48361Y}{1344}{e}^2\right)\pi v^5
\right. \cr && \left.\hspace*{20mm}
-\left(\frac{2633}{224}-\frac{4301Y^2}{224}
+\left\{\frac{66139}{1344}-\frac{18419Y^2}{448}\right\}{e}^2\right)q v^5
\right],\cr
&&\left<\frac{d\mathcal{C}}{d t}\right> =
-\frac{64}{5}\left(\frac{\mu}{M^2}\right) M^2 v^{6}(1-{e}^2)^{3/2}
\left({1-Y^2}\right)\left[
\left(1+\frac{7}{8}{e}^2\right)
\right. \cr && \left.\hspace*{20mm}
-\left(\frac{743}{336}-\frac{23}{42}{e}^2\right)v^2
-\left(\frac{85Y}{8}+\frac{211Y}{8}{e}^2\right)qv^3
\right. \cr && \left.\hspace*{20mm}
+\left(4+\frac{97}{8}{e}^2\right)\pi v^3
-\left(\frac{129193}{18144}+\frac{84035}{1728}{e}^2\right)v^4
\right. \cr && \left.\hspace*{20mm}
-\left(\frac{329}{96}-\frac{53Y^2}{8}+\left\{
\frac{929}{96}-\frac{163Y^2}{8}\right\}{e}^2\right)q^2 v^4
\right. \cr && \left.\hspace*{20mm}
+\left(\frac{2553Y}{224}-\frac{553Y}{192}{e}^2\right)qv^5
-\left(\frac{4159}{672}+\frac{21229}{1344}{e}^2\right)\pi v^5
\right].
\end{eqnarray}
Here, a factor of $(1-{e}^2)^{3/2}$ is factored out  
in order to make it easier to compare these results with the
well-known formulas derived by Peters and Mathews~\cite{PM}.

We can derive the evolution of the orbital parameters,
$\iota^i=\{v,e,Y\}$, from the rates of change of the integrals
of motion, $I^i=\{{\cal E},{\cal L},{\cal C}\}$,
\begin{equation}
\left<\frac{d\iota^i}{dt}\right>=(G^{-1})^i_j 
\left<\frac{dI^j}{dt}\right>; \qquad
G^i_j=\frac{\partial I^i}{\partial \iota^j}.
\end{equation}
Using the above relation with Eqs.~(\ref{eq:kekka_EL}) 
and (\ref{Ydef}), the evolutions of the parameters
$v,{e}$ and $Y$ are obtained as

\begin{eqnarray}
&& \left<\frac{dv}{dt}\right> =
  \frac{32}{5}\left(\frac{\mu}{M^2}\right)v^9(1-{e}^2)^{3/2}
\left[
\left(1+\frac{7}{8}{e}^2\right)
-\left(\frac{743}{336}+\frac{55}{21}{e}^2\right)v^2
\right. \cr && \left.\hspace*{20mm}
-\left(\frac{133Y}{12}+\frac{379Y}{24}{e}^2\right)qv^3
+\left(4+\frac{97}{8}{e}^2\right)\pi v^3
\right. \cr && \left.\hspace*{20mm}
+\left(\frac{34103}{18144}-\frac{526955}{12096}{e}^2\right)v^4
\right. \cr && \left.\hspace*{20mm}
-\left(\frac{329}{96}-\frac{815Y^2}{96}
+\left\{\frac{929}{96}-\frac{477Y^2}{32}\right\}{e}^2\right)q^2v^4
\right. \cr && \left.\hspace*{20mm}
-\left(\frac{1451Y}{56}+\frac{1043Y}{96}{e}^2\right)qv^5
-\left(\frac{4159}{672}+\frac{48809}{1344}{e}^2\right)\pi v^5
\right],\cr
&& \left< \frac{d{e}}{dt}\right>  =
-\frac{304}{15}\left(\frac{\mu}{M^2}\right)v^8{e}\,(1-{e}^2)^{3/2}
\left[
\left(1+\frac{121}{304}{e}^2\right)
-\left(\frac{6849}{2128}+\frac{4509}{2128}{e}^2\right)v^2
\right. \cr && \left.\hspace*{20mm}
-\left(\frac{879Y}{76}+\frac{515Y}{76}{e}^2\right)qv^3
+\left(\frac{985}{152}+\frac{5969}{608}{e}^2\right)\pi v^3
\right. \cr && \left.\hspace*{20mm}
-\left(\frac{286397}{38304}+\frac{2064415}{51072}{e}^2\right)v^4
\right. \cr && \left.\hspace*{20mm}
-\left(\frac{3179}{608}-\frac{5869Y^2}{608}
+\left\{\frac{8925}{1216}-\frac{10747Y^2}{1216}\right\}{e}^2\right)q^2v^4
\right. \cr && \left.\hspace*{20mm}
-\left(\frac{1903Y}{304}-\frac{22373Y}{8512}{e}^2\right)qv^5
-\left(\frac{87947}{4256}+\frac{4072433}{68096}{e}^2\right)\pi v^5
\right],\cr
&&\left< \frac{dY}{dt}\right> =
 -\frac{244}{15}\left(\frac{\mu}{M^2}\right)v^8 (1-{e}^2)^{3/2}(1-Y^2)
\left[
\left(1+\frac{189}{61}{e}^2\right)qv^3
\right. \cr && \left.\hspace*{20mm}
-\left(\frac{13Y}{244}+\frac{277Y}{244}{e}^2\right)q^2 v^4
-\left(\frac{10461}{1708}+\frac{83723}{3416}{e}^2\right)qv^5
\right].
\label{iotas}
\end{eqnarray}
Substituting $Y=1$, it is found that
these results are consistent with the previous results
~\cite{Glampedakis:2005hs, Gair, Tagoshi},
except for the errors reported in our previous 
paper~\cite{Sago:2006}.

%%%%%%%%%%%%%%%%%%%%%%%%%%%%%%%%%%%%%%%%%%%%%%%%%%%%%%%%%%%%%%%%%%%%%%
\subsection{Phase evolution of gravitational waves} \label{phase}
%%%%%%%%%%%%%%%%%%%%%%%%%%%%%%%%%%%%%%%%%%%%%%%%%%%%%%%%%%%%%%%%%%%%%%

The time dependence of the orbital parameters gives us information 
about the phase evolution of gravitational waves.
There are three fundamental frequencies,
$\omega_r=\Omega_r \left<dt/d\lambda\right>_{\lambda}^{-1}$,
$\omega_\theta=\Omega_\theta \left<dt/d\lambda\right>_{\lambda}^{-1}$,
and $\omega_\varphi=\left<d\varphi/d\lambda\right> 
 \left<dt/d\lambda\right>_{\lambda}^{-1}$.
$\omega_r$ is the angular velocity of radial oscillations,  
and it is not important for small eccentric orbits.
By contrast, $\omega_\theta$ and $\omega_\varphi$ are 
equally fundamental for greatly inclined orbits.
Stated briefly, gravitational waves cannot be expressed by a single phase,
even if we pick up the dominant quadrupole contribution. 
%Instead, they are expressed as $h(t)=h_\theta(t)+h_\varphi(t)$,
%with $h_\theta(t)=A_\theta(t)\cos(2\int \omega_\theta dt)$ 
%and $h_\varphi(t)=A_\varphi(t)\cos(2\int \omega_\varphi dt)$.

When the orbit is in the equatorial plane, $\omega_\varphi$ 
corresponds to the angular velocity used in 
the standard post-Newtonian calculation~\cite{PoissonWill:95}. 
%However, when the orbit has inclination, 
%both of them differ from the post-Newtonian frequency. 
For general orbits, the post-Newtonian angular velocity is given 
by $\omega_N\equiv (1-Y) \omega_\theta+ Y\omega_\varphi$. 
This can be understood in the following way. 
First we consider the frame rotating 
at the angular velocity of the orbital precession due to the 
spin-orbit coupling, $\omega_\theta -\omega_\varphi$. 
In this frame, the orbital plane of the geodesic motion is fixed. 
The orbital angular velocity observed in this frame is
given by $\omega_\theta$. 
Then, the effect of the rotating frame 
can be understood as the sum of two rotations,
that around the axis perpendicular to the orbital 
plane and that around an axis in the orbital plane. 
The first rotation, whose angular velocity is 
$Y(\omega_\theta -\omega_\varphi)$, is understood
as the relative motion between the rotating frame
and the non-rotating one.
Therefore, the angular velocity of the orbit projected onto the 
momentary orbital plane is given 
by $\omega_N=(1-Y) \omega_\theta+ Y\omega_\varphi$. 
This is in some sense the angular velocity observed from the 
direction perpendicular to the orbital plane. 
The latter rotation causes the change of the orbital plane. 
As the orbital plane itself changes, this post-Newtonian 
angular velocity $\omega_N$ does not appear in the waveform with 
a fixed direction of the observer.  
When we are concerned with a short period of time in which the 
precession of the orbital plane can be ignored, of course, 
there is no practical way to distinguish these three angular velocities, 
$\omega_N$, $\omega_\varphi$ and $\omega_\theta$. 

In this subsection, we consider the evolution of 
the frequency 
$F_\varphi\equiv \omega_\varphi/\pi$ 
and the corresponding phase 
$\Phi_\varphi(F_{\varphi})\equiv 2\pi\! \int\! F_\varphi dt$, 
assuming that the evolution of all orbital parameters are 
governed by the time averaged expressions %!!! +s
given in Eqs.~(\ref{iotas}).
Here, we briefly describe the method for obtaining analytic expressions for them.
The details, including the derivation of 
the Fourier transformed waveform under the stationary phase 
approximation and the results for $F_\theta\equiv \omega_\theta/\pi$,
are reported in Appendix B. 

Using Eq.~(\ref{2.49}), we can express $F_\varphi$ 
in terms of $\iota^j$. Solving this relation inversely with respect 
to $v$, we can express $v$ as a function of $F_{\varphi}$, $e$ and $Y$ 
in the post-Newtonian expansion, i.e. expansion in powers of  
$F_{\varphi}$. Then, taking the derivative of $F_{\varphi}(\iota^j)$, 
we rewrite it using Eqs.~(\ref{iotas}). %!!! +()
Substituting $v(F_\varphi,e,Y)$ 
into this expression, we obtain  
\begin{eqnarray}
\frac{dF_\varphi}{dt}&=&
\frac{96}{5\pi\mathcal{M}^2}
(\pi\mathcal{M}F_\varphi)^\frac{11}{3}
\left[\left(1+\frac{157}{24}e^2\right)
-\left(\frac{743}{336}+\frac{3683}{112}e^2\right)\pMFp^\frac{2}{3}
\right. \cr &&{}\left.
-\left(\frac{10}{3}+\frac{73Y}{12}
+\left\{\frac{1193}{36}+\frac{857Y}{24}\right\}e^2\right)q\pMFp
\right. \cr && {}\left.
+\left(4+\frac{2335}{48}e^2\right)\pi\pMFp
+\left(\frac{34103}{18144}-\frac{89353}{1728}e^2 \right)\pMFp^\frac{4}{3}
\right. \cr && {}\left.
+\left(-\frac{233}{96}+2Y+\frac{527}{96}Y^2
+\left\{-\frac{2875}{96}+\frac{281}{12}Y+\frac{4165}{96}Y^2 \right\} e^2
\right)q^2\pMFp^\frac{4}{3}
\right. \cr && {}\left.
-\left(\frac{4159}{672}+\frac{20135}{96}e^2\right)\pi \pMFp^\frac{5}{3}
\right. \cr && {}\left.
+\left(\frac{743}{72}-\frac{13907Y}{336}
+\left\{\frac{16687}{84}-\frac{84365Y}{336} \right\}e^2
\right)q \pMFp^\frac{5}{3}
\right],
\label{dFvarphidt} 
\end{eqnarray}
where we have introduced 
the chirp mass, $\mathcal{M}\equiv M^{2/5}\mu^{3/5}$. 
The evolution equation for $e$ and $Y$ can also be rewritten 
by using $F_\varphi$ instead of $v$. 
We have to integrate these three differential equations simultaneously. 
This may appear to be difficult, but 
we can perform this integration iteratively,
assuming a small eccentricity and the post-Newtonian expansion. 
It is important to note that 
$\delta Y\equiv Y-Y_I$ is 
also small in the post-Newtonian expansion, where $Y_I$ is the value of
$Y$ at $F_\varphi=0$. 
For this reason, in the lowest order approximation, 
we can fix the value of $Y$ to $Y_I$. 
The effect of the variation $\delta Y$ can be taken into account 
perturbatively. 
Eliminating $dt$ from these differential equations, 
$e^2$ and $Y$ are integrated as functions of $F_\varphi$ as 

\begin{eqnarray}
 e^2 & = & \tilde e_I^2 (\pi M F_{\varphi})^{-{19\over 9}}
     \Biggl(1+{3215\over 1008}\pMFp^{2\over 3}
\cr &&
        +\left\{-{377\over 72}\pi+q\left({38\over 9}
        -{287\over 54}Y_I\right)\right\}\pMFp\cr
&& +\left\{{35705555\over 1524096}+q^2\left(
            {731\over 576}-{19\over 6}Y_I+{1675\over 576}Y_I^2\right)
             \right\}\pMFp^{4\over 3}\cr
&& +\left\{-{269971\over 25920}\pi+q\left(
            {41795\over 4536}-{5435639\over 272160}Y_I\right)
             \right\}\pMFp^{5\over 3}+\cdots + 
   \delta e^2_{(\delta Y)}\Biggr),\cr
\delta e^2_{(\delta Y)}& = &
       -{17507q^2\over 432}(1-Y_I^2)
        \pMFp^2+\cdots, 
\cr
 \delta Y & = & - (1-Y_I^2)q \pMFp \left(
             {61\over 72}
            -{163\over 384}\tilde e_I^2 \pMFp^{-{19\over 9}}
                +\cdots\right).
\label{eq:eY-evolve}
\end{eqnarray} 
Here, $\tilde e_I$ denotes the limiting value of 
$e\times (\pi M F_\varphi)^{19/18}$ in the limit $F_\varphi\to 0$. 
We have denoted the leading-order correction to ${e}^2$ due to 
the evolution of $Y$ separately by $\delta e^2_{(\delta Y)}$,   
although it is relatively order 3PN.

Substituting the relations in Eq.(\ref{eq:eY-evolve})
into Eq.(\ref{dFvarphidt})
and integrating $F_\varphi (dF_\varphi/dt)^{-1}$
over $F_\varphi$,
the phase $\Phi_\varphi(F_\varphi)$ is  
calculated as

\begin{eqnarray}
\Phi_\varphi(F_\varphi)&=&\phi_\varphi^{\,c}-\frac{1}{16}
(\pi \mathcal{M} F_\varphi)^{-\frac{5}{3}}
\left[1+\frac{3715}{1008}\pMFp^\frac{2}{3}
\right. \cr && \left.{}
+\left\{q\left(\frac{25}{3}+\frac{365}{24}Y_I\right)-10\pi\right\} \pMFp 
\right. \cr && \left.
{}+ \left\{ \frac{15293365}{1016064}
+q^2\left(\frac{1165}{96}-10Y_I-\frac{2635}{96}Y_I^2\right) \right\}
\pMFp^\frac{4}{3}
\right. \cr && \left.
{}+\left\{ \frac{38645}{672}\pi 
- q\left(\frac{3715}{168}+\frac{688405}{2016}Y_I\right) \right\}
\pMFp^\frac{5}{3}\ln\pMFp 
\right. \cr && \left. {}
+\delta\Phi_{\varphi}^{(\delta Y)} 
+\delta\Phi_{\varphi}^{(e)}
\right],
\cr \cr
\delta\Phi_\varphi^{(\delta Y)}&=&q^2 (1-Y_I^2) 
\left(
\frac{22265}{864}
+\frac{21072205}{73728} \tilde e_I^2 \pMFp^{-\frac{19}{9}}\right)
\pMFp^2,
\cr \cr
\delta\Phi_\varphi^{(e)}&=&\tilde e_I^2 \pMFp^{-\frac{19}{9}}
\left[-\frac{785}{272}-\frac{2045665}{225792}\pMFp^\frac{2}{3}
\right. \cr && \left.
{}+\left\{ \frac{65561}{2880}\pi 
-q\left(\frac{2057}{90}+\frac{2953}{540}Y_I \right)\right\} \pMFp
\right. \cr && \left.
{}-\left\{\frac{111064865}{10948608}+
\left(-\frac{698695}{101376}+\frac{16895}{1056}Y_I
+\frac{650665}{101376}Y_I^2\right) q^2 \right\} \pMFp^\frac{4}{3}
\right. \cr && \left.
{}+\left\{\frac{3873451}{86184}\pi-q\left(
\frac{1247185}{24624}+\frac{1899015067}{8273664}Y_I\right)\right\}
\pMFp^{\frac{5}{3}} \right],
\label{Phivarphi}
\end{eqnarray}
where $\phi_\varphi^{\,c}$ is a constant of integration, and
$\delta\Phi_\varphi^{(e)}$ and $\delta\Phi_\varphi^{(\delta Y)}$ 
express the corrections of $O(\tilde e_I^2)$ and 
the terms associated with the time variation of the 
inclination angle, respectively. 
We have included the cross terms between 
these two effects in $\delta\Phi_\varphi^{(\delta Y)}$. 

Here, several remarks are in order. 
The terms of $O(\tilde e_I^2)$ seem to be large
in the sense of the post-Newtonian order. 
Those terms have relatively large inverse powers of $F_\varphi$. 
However, we should recall that 
$e^2\approx \tilde e_I^2 \pMFt^{-19/9}$. 
Therefore, under the current assumption of a small eccentricity 
$(e\ll 1)$, the terms associated with the factor 
$\tilde e_I^2 \pMFt^{-19/9}$ 
are %!!! is -> are
much smaller than the other terms. 
In the above calculation, we have basically kept the 
terms up through $O(v^5, e^2)$.
However, for $\delta\Phi_\varphi^{(\delta Y)}$, 
we have also kept the terms of higher post-Newtonian order, $O(v^6)$,  
since the leading-order correction due to $\delta Y$ 
starts with this order. 

As mentioned above, more explicit formulas including 
the Fourier transform of the waveform under
the stationary phase approximation are given in Appendix B. 
Here, we would like to stress that the analytic method for computing
the phase evolution presented here 
can be extended systematically to higher order in $v$ and $e$. 

%%%%%%%%%%%%%%%%%%%%%%%%%%%%%%%%%%%%%%%%%%%%%%%%%%%%%%%%%%%%%%%%%%%%%%
\section{Summary} \label{summary}
%%%%%%%%%%%%%%%%%%%%%%%%%%%%%%%%%%%%%%%%%%%%%%%%%%%%%%%%%%%%%%%%%%%%%%

In this paper, we have improved an analytic method 
for calculating the adiabatic evolution of the orbital 
parameters of a point particle orbiting 
a Kerr black hole. We have removed 
the previous limitation to an orbit with a small inclination 
angle. To do this, we rewrote the source term of the Teukolsky equation, 
removing inverse powers of $\sin\theta_z$ by using integration by parts, 
where $\theta_z$ is the polar angle of the particle position. 
As a result, our new expression for the source term is much more 
complicated than the original one, but it is possible to carry out an
analytic Fourier transform 
of this source term. Moreover, even in a numerical 
approach along the line of Refs.~\citen{drasco} and \citen{fujita}, 
our expression might be useful when we consider the polar 
orbit, since the singular behavior of the source term near the poles 
in the original formulation is completely removed.  

We have also shown that it is possible to integrate the time 
evolution of the orbital parameters under the assumption that 
it is governed by the time averaged values derived in the
adiabatic approximation. Furthermore, analytic calculation 
of the Fourier transform of the waveform is also demonstrated 
in Appendix B. 
The results presented here are restricted 
to $O(v^5, e^2)$, where $v$ and $e$ are, respectively, 
the velocity and the eccentricity of the particle.  
Our method, however, can be extended systematically to higher 
order. Such analytic expressions for the waveform should be 
useful for the fast generation of templates of 
gravitational waveforms. 

Our analytic expressions for the 
averaged evolutions of orbital parameters (\ref{iotas}) 
are written as polynomial functions with respect to $Y$ 
defined by (\ref{Ydef}). This result could be anticipated 
from the expressions for orbits given in \S\ref{Orbits}. 
The only non-polynomial expression is 
$\sqrt{1-Y^2}$ in $\cos\theta(\lambda)$. 
However, changing the signature of $\cos\theta(\lambda)$ alone 
corresponds to simply flipping the direction of the $z$-axis. 
This change should not 
affect the rates of change of the energy, the angular momentum 
and the Carter constant, which means that only 
even powers of $\cos\theta(\lambda)$ contribute to 
these rates of change. Here we do not pursue a rigorous proof of 
this statement. 
If we can prove that the averaged evolutions of orbital parameters can
be written as polynomials in Y,
we will not have to perform computations for a large inclination angle, 
at least for analytic calculations, because
the expressions obtained 
in the expansion of a small inclination angle will be sufficient 
to determine all the coefficients that we want to know. 
Then, extrapolation to a large inclination angle is exact.  
To do this, we also need to know the maximum 
power of $Y$ at each post-Newtonian order. This issue will be 
studied in a future publication. 

\section*{Acknowledgements}
We would like to thank T.~Nakamura, M.~Sasaki and H.~Tagoshi
for useful discussions and comments. 
HN, KG and WH are also grateful to all the participants in the
Capra 9 Meeting \& Workshop at the University of Wisconsin-Milwaukee 
for invaluable discussions. 
%the University of Texas Brownsville 
%and HN, NS and WH thank all the participants of 
%8th Capra Meeting on Radiation Reaction at the Rutherford Appleton Laboratory 
WH is supported by the JSPS Research Fellowships 
for Young Scientists, No.~1756.
HN is supported by JSPS Postdoctoral Fellowships 
for Research Abroad and in part by 
the NSF for financial support from grant PHY-0354867.
NS's work was supported by PPARC through grant number PP/D001110/1.
This work is supported in part 
by Grant-in-Aid for Scientific Research, Nos.
16740141 and 17340075, 
and by the 21st Century COE
``Center for Diversity and Universality in Physics'' at 
Kyoto University, both from the Ministry of
Education, Culture, Sports, Science and Technology of Japan.

\appendix

%%%%%%%%%%%%%%%%%%%%%%%%%%%%%%%%%%%%%%%%%%%%%%%%%%%%%%%%%%%%%%%%%%%%%%
\section{Reformulation of the Teukolsky equation for $|m|\le 1$}
\label{app:mle1}
%%%%%%%%%%%%%%%%%%%%%%%%%%%%%%%%%%%%%%%%%%%%%%%%%%%%%%%%%%%%%%%%%%%%%%

\begin{itemize}
\item $m=0$. In this case, 
${}_\sigma\tilde{S}=\sin^{2\sigma}\theta \times$ 
  (finite polynomial of $\cos\theta$)
in the PN approximation. Thus, we can write Eq.~(\ref{reformalized Base}) as
\begin{eqnarray}
 Z_{\Lambda}&=&\frac{\mu}{2i\omega B^{\rm{inc}}}
 \int_{-\infty}^{\infty} d\lambda
 e^{i\omega t-im(\left<\frac{d\varphi}{d\lambda}\right>_{\!\!\lambda} 
  +\varphi^{(r)})}
 \left( \frac{-1}{2\sqrt{2\pi}}\frac{\bar{\rho}}{\rho} \right)
 \Bigg[ \frac{D_r^2}{\Delta^2}
 \left({}_0\Xi_m -2ia\rho\, {}_1\Xi_m \right) R_{\Lambda} \cr
 &-&\frac{2D_r (D_\theta \sin^2\theta)}{\Delta}
 \left(\frac{{}_1\Xi_m}{\sin^2\theta}(\mathcal{J}_{-}-(\rho+\bar{\rho}))
 +ia\frac{{}_2\Xi_m}{\sin^2\theta}
 (\bar{\rho}-\rho)\mathcal{J}_{-}\right) R_{\Lambda} \cr
 &+&(D_\theta \sin^2\theta)^2 \frac{{}_2\Xi_m}{\sin^4\theta}
 (\mathcal{J}_{-}^{2}-2\rho\mathcal{J}_{-}) 
 R_{\Lambda} \Bigg]_{r=r(\lambda)}.
\end{eqnarray}
This form has no factors of $1/\sin\theta$.

\item $m=\pm 1$. In this case, the $D_\theta \, {}_1\Xi_m$ and
$D_\theta^2\,{}_2\Xi_m$ terms have $1/\sin^2\theta$ factors.
We can treat the $D_\theta\,{}_1\Xi_m$ term here in a manner
similar to
that for $|m|\ge 2$. Therefore, we have to consider the
$D_\theta^2\,{}_2\Xi_m$ term.
Only in the case $m=\pm 1$, we can find the equation
\begin{eqnarray}
 \frac{d}{d\lambda} X_{\pm 1}=\pm \left[
  D_\theta+\left(\frac{1}{1\pm \cos\theta}\frac{d\cos\theta}{d\lambda}
  -ia\mathcal{E}
  +i\left<\frac{\mathcal{L}}{\sin^2\theta}\right>_{\!\!\lambda} \right)
 \right]X_{\pm 1}.
\end{eqnarray}
Since ${}_2\tilde{S}$ is proportional to $\sin^2\theta(1\pm \cos\theta)$,
we obtain the relation
\begin{eqnarray}
 D_\theta^2 \, {}_2\Xi_{\pm 1}&=&(D_\theta \sin^2\theta)
 \frac{{}_2\tilde{S}}{\sin^2\theta(1\pm\cos\theta)}
 \left[
  -\frac{d\cos\theta}{d\lambda}
  \right. \cr && \left. 
  +(1\pm\cos\theta)
  \left(\pm\frac{d}{d\lambda}+ia\mathcal{E}
   -i\left<\frac{\mathcal{L}}{\sin^2\theta}\right>_{\!\!\lambda} \right)
 \right] X_{\pm 1}.
\end{eqnarray}
Therefore, the amplitude of the partial wave for $m=\pm1$ is %!!! -as
      %follows,
\begin{eqnarray}
 Z_{\Lambda}&=&\frac{\mu}{2i\omega B^{\rm{inc}}}
 \int_{-\infty}^{\infty} d\lambda \,
 e^{i\omega t-im(\left<\frac{d\varphi}{d\lambda}\right>_{\!\!\lambda} 
  +\varphi^{(r)})}
 \left( \frac{-1}{2\sqrt{2\pi}}\frac{\bar{\rho}}{\rho} \right)
 \cr && \times 
 \Bigg[ \frac{D_r^2}{\Delta^2}
 \left({}_0\Xi_m -2ia\rho\,{}_1\Xi_m \right) R_{\Lambda} 
 -\frac{2D_r}{\Delta}
 \biggl\{ F_m({}_0\Xi_m,{}_1\Xi_m)(\mathcal{J}_{-}-(\rho+\bar{\rho}))
  \cr && \quad
 +ia(D_\theta \sin^2\theta)\frac{{}_2\Xi_m}{\sin^2\theta}
 (\bar{\rho}-\rho)\mathcal{J}_{-}\biggr\} R_{\Lambda} 
 \cr && \quad
 + \frac{(D_\theta \sin^2\theta) \, {}_2\tilde{S}}
 {\sin^2\theta(1+m\cos\theta)}
 \left[-\frac{d\cos\theta}{d\lambda}X_m
 \right. \cr && \left. \quad 
 +(1+m\cos\theta)
  \left(m\frac{d}{d\lambda}+ia\mathcal{E}
  -i\left<\frac{\mathcal{L}}{\sin^2\theta}\right>_{\!\!\lambda} 
   \right)X_m\right]
 \cr && \quad \times 
 (\mathcal{J}_{-}^{2}-2\rho\mathcal{J}_{-})
 R_{\Lambda} \Bigg]_{r=r(\lambda)}.
\end{eqnarray}
\end{itemize}

\section{Formulas for phase evolution}

In this appendix we present all the detailed formulas 
for the orbital and phase evolutions to supplement \S\ref{phase}. 
The expression for ${dF_\theta}/{dt}$ corresponding to 
(\ref{dFvarphidt}) is
\begin{eqnarray}
\frac{dF_\theta}{dt}&=&
\frac{96}{5\pi\mathcal{M}^2}(\pi\mathcal{M}F_\theta)^\frac{11}{3}
\left[\left(1+\frac{157}{24}e^2\right)
-\left(\frac{743}{336}+\frac{3683}{112}e^2\right)\pMFt^\frac{2}{3}
\right. \cr &&{}\left.
-\left(\frac{73Y}{12}+\frac{857Y}{24}e^2\right)q\pMFt
\right. \cr && {}\left.
+\left(4+\frac{2335}{48}e^2\right)\pi\pMFt
+\left(\frac{34103}{18144}-\frac{89353}{1728}e^2 \right)\pMFt^\frac{4}{3}
\right. \cr && {}\left.
+\left(-\frac{233}{96}+\frac{527}{96}Y^2
+\left\{-\frac{2875}{96}+\frac{4165}{96}Y^2 \right\} e^2
\right)q^2\pMFt^\frac{4}{3}
\right. \cr && {}\left.
-\left(\frac{4159}{672}+\frac{20135}{96}e^2\right)\pi \pMFt^\frac{5}{3}
\right. \cr && {}\left.
-\left(\frac{13907Y}{336}+\frac{84365Y}{336}e^2
\right)q \pMFt^\frac{5}{3}
\right].
\end{eqnarray}
As mentioned in \S\ref{phase}, the standard post-Newtonian 
calculation uses a frequency that is defined differetly, namely
$F_N=(1-Y) F_\theta+ YF_\varphi$.
For comparison, we also give an expression for $dF_N/dt$,
\begin{eqnarray}
\frac{dF_N}{dt}&=&
\frac{96}{5\pi\mathcal{M}^2}(\pi\mathcal{M}F_N)^\frac{11}{3}
\left[1-\frac{743}{336}\pMFN^\frac{2}{3}
+\left(4\pi-\frac{113Y}{12}q\right) \pMFN
\right. \cr && {}\left.
+\left\{\frac{34103}{18144}
+\left(-\frac{233}{96}+2Y+\frac{527}{96}Y^2
\right)q^2\right\}\pMFN^\frac{4}{3}
\right].
\end{eqnarray}
This agrees with the standard post-Newtonian result, up to 
1.5PN~\cite{PoissonWill:95}.

The expression for the phase evolution of the $\theta$-oscillation
corresponding to (\ref{Phivarphi}) is
\begin{eqnarray}
\Phi_\theta(F_\theta)&=&\phi_\theta^{\,c}-\frac{1}{16}
(\pi \mathcal{M} F_\theta)^{-\frac{5}{3}}
\left[1+\frac{3715}{1008}\pMFt^\frac{2}{3}
+\left\{\frac{365}{24}Y_Iq-10\pi\right\} \pMFt 
\right. \cr && \left.
{}+ \left\{ \frac{15293365}{1016064}
+q^2\left(\frac{1165}{96}-\frac{2635}{96}Y_I^2\right) \right\}
\pMFt^\frac{4}{3}
\right. \cr && \left.
{}+\left\{ \frac{38645}{672}\pi 
- q\frac{688405}{2016}Y_I \right\}
\pMFt^\frac{5}{3}\ln\pMFt +\delta\Phi_{\theta}^{(\delta Y)} 
+\delta\Phi_{\theta}^{(e)}
\right],
\cr \cr
\delta\Phi_\theta^{(\delta Y)}&=&q^2 (1-Y_I^2) 
\left(
\frac{22265}{864}
+\frac{21072205}{73728} e_I^2 \pMFt^{-\frac{19}{9}}\right)
\pMFt^2,
\cr \cr
\delta\Phi_\theta^{(e)}&=&e_I^2 \pMFt^{-\frac{19}{9}}
\left[-\frac{785}{272}-\frac{2045665}{225792}\pMFt^\frac{2}{3}
\right. \cr && \left.
{}+\left\{ \frac{65561}{2880}\pi 
-q\frac{2953}{540}Y_I \right\} \pMFt
\right. \cr && \left.
{}-\left\{\frac{111064865}{10948608}+
\left(\frac{698695}{101376}
-\frac{650665}{101376}Y_I^2\right) q^2 \right\} \pMFt^\frac{4}{3}
\right. \cr && \left.
{}+\left\{\frac{3873451}{86184}\pi-q
\frac{1899015067}{8273664}Y_I\right\}
\pMFt^{\frac{5}{3}} \right],
\end{eqnarray}
where $\phi_\theta^{\,c}$ is a constant of integration.

In order to compute the waveform, we also need to know 
$t(F)\equiv\int (dF/dt)^{-1} dF$. 
This computation can be carried out
in a manenr similar to that for the calculation of the phase $\Phi(F)$. 
The results are 

\begin{eqnarray}
t_\varphi(F_\varphi)&=&t_\varphi^{\,c}-\frac{5}{256}\mathcal{M}
(\pi \mathcal{M} F_\varphi)^{-\frac{8}{3}}\left[
1+\frac{743}{252}\pMFp^\frac{2}{3}
\right. \cr && \left.
{}
+\left\{q\left(\frac{16}{5}+\frac{146}{15}Y_I\right)-\frac{32}{5}\pi
\right\}\pMFp
\right. \cr && \left.
{}+\left\{\frac{3058673}{508032}+\left(\frac{233}{48}-4Y_I
-\frac{527}{48}Y_I^2 \right)q^2\right\} \pMFp^\frac{4}{3}
\right. \cr && \left.
{}+\left\{q\left(\frac{743}{63}+\frac{137681}{756}Y_I\right)
-\frac{7729}{252}\pi\right\}
\pMFp^{\frac{5}{3}}
 + \delta t_\varphi^{(\delta Y)} 
+ \delta t_\varphi^{(e)} \right],
\cr \cr
\delta t_\varphi^{(\delta Y)}&=&-q^2(1-Y_I^2)
\left(
\frac{4453}{216}
-\frac{4214441}{14400}e_I^2\pMFp^{-\frac{19}{9}}\right)
\pMFp^2,
\cr \cr
\delta t_\varphi^{(e)}&=&e_I^2 \pMFp^{-\frac{19}{9}}\left[
-\frac{157}{43}-\frac{409133}{37296}\pMFp^\frac{2}{3}
\right. \cr && \left.
{}
+\left\{\frac{65561}{2448}\pi-\left(\frac{242}{9}+\frac{2953}{459}Y_I
\right)\right\} \pMFp
\right. \cr && \left.
{}+\left\{-\frac{22212973}{1928448}+\left(-\frac{139739}{17856}
+\frac{109}{6}Y_I+\frac{130133}{17856}Y_I^2\right)q^2\right\}\pMFp^\frac{4}{3}
\right. \cr && \left.
{}+\left\{\frac{3873451}{79380}\pi
-\left(\frac{249437}{4536}+\frac{1899015067}{7620480}Y_I\right)
\right\}\pMFp^\frac{5}{3}\right],
\end{eqnarray}
\begin{eqnarray}
t_\theta(F_\theta)&=&t_\theta^{\,c}-\frac{5}{256}\mathcal{M}
(\pi \mathcal{M} F_\theta)^{-\frac{8}{3}}\left[
1+\frac{743}{252}\pMFt^\frac{2}{3}
+\left\{q\frac{146}{15}Y_I-\frac{32}{5}\pi
\right\}\pMFt
\right. \cr && \left.
{}+\left\{\frac{3058673}{508032}+\left(\frac{233}{48}
-\frac{527}{48}Y_I^2 \right)q^2\right\} \pMFt^\frac{4}{3}
\right. \cr && \left.
{}+\left\{q\frac{137681}{756}Y_I-\frac{7729}{252}\pi \right\}
\pMFt^{\frac{5}{3}}
+ \delta t_\theta^{(\delta Y)} 
+ \delta t_\theta^{(e)} 
\right],
\cr \cr
\delta t_\theta^{(\delta Y)}&=&-q^2(1-Y_I^2)
\left(
\frac{4453}{216}
-\frac{4214441}{14400}e_I^2\pMFt^{-\frac{19}{9}}\right)
\pMFt^2,
\cr \cr
\delta t_\theta^{(e)}&=&e_I^2 \pMFt^{-\frac{19}{9}}\left[
-\frac{157}{43}-\frac{409133}{37296}\pMFt^\frac{2}{3}
\right. \cr && \left.
{}
+\left\{\frac{65561}{2448}\pi-\frac{2953}{459}qY_I\right\} \pMFt
\right. \cr && \left.
{}+\left\{-\frac{22212973}{1928448}+\left(-\frac{139739}{17856}
+\frac{130133}{17856}Y_I^2\right)q^2\right\}\pMFt^\frac{4}{3}
\right. \cr && \left.
{}+\left\{\frac{3873451}{79380}\pi
-\frac{1899015067}{7620480}qY_I
\right\}\pMFt^\frac{5}{3}\right],
\end{eqnarray}
where $t_\varphi^{\,c}$ and $t_\theta^{\,c}$ are constants of integration.

Writing the Fourier components 
of the waveforms as 
$\tilde{h}(f)=\mathcal{A} e^{i\psi(f)}$, 
$\psi(f)$ is calculated as 
$\psi(f)=-\Phi(f)+2\pi f t(f)$, 
using the stationary phase approximation~\cite{spa}. 
The final results are
\begin{eqnarray}
\psi_\varphi(f)&=&2\pi ft_\varphi^{\,c}-\phi_\varphi^{\,c}
+\frac{3}{128}\pMft^{-\frac{5}{3}}
\left[1+\frac{3715}{756}\pMft^\frac{2}{3}
\right. \cr && \left.
{}+\left\{\frac{73}{3}Y_Iq -16\pi\right\}\pMft
\right. \cr && \left.
{}+\left\{\frac{15293365}{508032}+q^2\left(\frac{1165}{48}
-\frac{2635}{48}Y_I^2\right)\right\}\pMft^\frac{4}{3}
\right. \cr && \left.
{}+\left\{\frac{38645}{252}\pi-\frac{688405}{756}qY_I
\right\}\pMft^\frac{5}{3} \ln\pMft
+\delta\psi_\varphi^{(\delta Y)}
+\delta\psi_\varphi^{(e)}
\right],
\cr\cr
\delta\psi_\varphi^{(\delta Y)}&=& q^2(1-Y_I^2)
\left(
\frac{111325}{1296}
-\frac{71645497}{138240}e_I^2\pMft^{-\frac{19}{9}}\right)
\pMft^2,
\cr\cr
\delta\psi_\varphi^{(e)}&=&e_I^2 \pMft^{-\frac{19}{9}}\left[
-\frac{2355}{1462}-\frac{2045665}{348096}\pMft^\frac{2}{3}
\right. \cr && \left.
{}+\left\{\frac{65561}{4080}\pi-\frac{2953}{765}Y_Iq\right\}\pMft
\right. \cr && \left.
{}+\left\{-\frac{111064865}{14141952}+q^2\left(-\frac{698695}{130944}
+\frac{650665}{130944}Y_I^2\right)\right\}
\pMft^\frac{4}{3}
\right. \cr && \left.
{}+\left\{\frac{3873451}{100548}\pi
-\frac{1899015067}{9652608}Y_Iq
\right\}\pMft^\frac{5}{3} \right],
\end{eqnarray}
and
\begin{eqnarray}
\psi_\theta(f)&=&\psi_\varphi(f)+2\pi f\delta t^{\,c}
 -\delta\phi^{\,c}\cr
&&
+\frac{3}{128}\pMft^{-\frac{5}{3}}
\left[\frac{40}{3}q\pMft
 -20Y_I q^2\pMft^\frac{4}{3}
\right. \cr && \left. {}
 -\frac{3715}{189}q \pMft^\frac{5}{3} \ln\pMft
+\delta\psi_\theta^{(e)}
\right],
\cr\cr
\delta\psi_\theta^{(e)}&=&e_I^2 \pMft^{-\frac{19}{9}}\left[
-\frac{242}{15}q\pMft
+\frac{545}{44}q^2 Y_I\pMft^\frac{4}{3}
\right. \cr && \left. {}
-\frac{1247185}{28728}q\pMft^\frac{5}{3} \right]. 
\end{eqnarray}
Here, we define $\delta t^{\,c}\equiv t_\theta^{\,c}-t_\varphi^{\,c}$, and 
we have absorbed a constant phase into $\phi_\varphi^{\,c}$ and 
$\delta\phi^{\,c}$.

\end{document}